\documentclass[aps,prb,amsmath,amssymb,twocolumn,superscriptaddress,preprintnumbers,showpacs,intlimits,longbibliography]{revtex4-2}
\pdfoutput=1
\bibpunct{[}{]}{,}{n}{}{}

\usepackage[utf8]{inputenc}
\usepackage{bm,latexsym,mathrsfs,enumerate}
\usepackage[dvipsnames]{xcolor}
\usepackage{mathtools}
\usepackage{makecell}
\usepackage{multirow}
\usepackage[breaklinks=true,unicode=true,urlcolor = blue,colorlinks = true,citecolor = blue,linkcolor = blue]{hyperref}
\usepackage{graphicx}
\usepackage{ragged2e}
\usepackage[export]{adjustbox}

\renewcommand{\vec}[1]{\bm{#1}}
\usepackage[final]{pdfpages}
\usepackage{tikz}
\makeatletter
\AtBeginDocument{\let\LS@rot\@undefined}
\makeatother
\renewcommand{\figurename}{\textbf{Figure}}

\begin{document}

\title{{\sffamily Direct evidence of a charge depletion region at the interface of Van der Waals monolayers and dielectric oxides: The case of superconducting FeSe/STO}
}

\author{\sffamily Khalil Zakeri}
\email{khalil.zakeri@kit.edu}

\affiliation{\sffamily Heisenberg Spin-dynamics Group, Physikalisches Institut, Karlsruhe Institute of Technology, Wolfgang-Gaede-Str. 1, D-76131 Karlsruhe, Germany}

\author{\sffamily Dominik Rau} \affiliation {\sffamily Heisenberg Spin-dynamics Group, Physikalisches Institut, Karlsruhe Institute of Technology, Wolfgang-Gaede-Str. 1, D-76131 Karlsruhe, Germany}

\author{\sffamily Janek Wettstein} \affiliation {\sffamily Heisenberg Spin-dynamics Group, Physikalisches Institut, Karlsruhe Institute of Technology, Wolfgang-Gaede-Str. 1, D-76131 Karlsruhe, Germany}

\author{\sffamily Markus D\"ottling} \affiliation {\sffamily Heisenberg Spin-dynamics Group, Physikalisches Institut, Karlsruhe Institute of Technology, Wolfgang-Gaede-Str. 1, D-76131 Karlsruhe, Germany}

\author{\sffamily Jasmin Jandke} \affiliation {\sffamily Physikalisches Institut, Karlsruhe Institute of Technology, Wolfgang-Gaede-Str. 1, D-76131 Karlsruhe, Germany}

\author{\sffamily Fang Yang} \affiliation {\sffamily Physikalisches Institut, Karlsruhe Institute of Technology, Wolfgang-Gaede-Str. 1, D-76131 Karlsruhe, Germany}

\author{\sffamily Wulf Wulfhekel} \affiliation {\sffamily Physikalisches Institut, Karlsruhe Institute of Technology, Wolfgang-Gaede-Str. 1, D-76131 Karlsruhe, Germany}

\affiliation {\sffamily Institute for Quantum Materials and Technologies, Karlsruhe Institute of Technology, D-76344, Eggenstein-Leopoldshafen, Germany}

\author{\sffamily J\"org Schmalian}

\affiliation {\sffamily Institute for Theory of Condensed Matter, Karlsruhe Institute of Technology, D-76131, Karlsruhe, Germany}

\affiliation {\sffamily Institute for Quantum Materials and Technologies, Karlsruhe Institute of Technology, D-76344, Eggenstein-Leopoldshafen, Germany}

\begin{abstract}

\end{abstract}
\maketitle

\onecolumngrid

 {\sffamily\textbf{\large{Abstract}}}\\
-----------------------------------------------------------------------------------------------------------------------------------------------------\\

 {\sffamily \textbf{The discovery of two dimensional Van der Waals materials has opened up several possibilities for designing novel devices. Yet a more promising way of designing exotic heterostutures with improved physical properties is to grow a monolayer of these materials on a substrate. For example, in the field of superconductivity it has been demonstrated that the superconducting transition temperature of a monolayer of FeSe grown on some oxide substrates e.g., strontium titanate (STO) is by far higher than its bulk counterpart. Although the system has been considered as a model system for understanding the phenomenon of high-temperature superconductivity, the physical mechanism responsible for this high transition temperature is still highly under debate. Here using momentum and energy resolved high-resolution electron energy-loss spectroscopy we probe the dynamic charge response of the FeSe/STO(001) system and demonstrate that the frequency- and momentum-dependent dynamic charge response is not compatible with a simple film/substrate model. Our analysis reveals the existence of a depletion region at the interface between this Van der Waals monolayer and the substrate. The presence of the depletion layer, accompanied with a considerably large charge transfer from STO into the FeSe monolayer, leads to a strong renormalization of the STO energy bands and a substantial band bending at the interface. Our results shed light on the electronic complexities of the FeSe/oxide interfaces and pave the way of designing novel low-dimensional high-temperature superconductors through interface engineering. We anticipate that the observed phenomenon is rather general and can take place in many two dimensional Van der Waals monolayers brought in contact with dielectric oxides or semiconducting substrates.}}\\
--------------------------------------------------------------------------------------------------------------------------------------------------------


\section{Introduction}\label{Sec:Intro}

Followed by the discovery of the unique physical properties of graphene, a monolayer (ML) of carbon atoms arranged in a honeycomb lattice, several other atomically thin two dimensional (2D) Van der Waals (VdW) materials have been discovered many of which exhibit unprecedented  electrical, optical and magnetic properties \cite{Geim2013, Novoselov2016,Jiang2021}. These materials can even be prepared in the form of a single layer in contact with a substrate. The physical properties of such hybrid structures can be precisely controlled by several means, opening up new opportunities for their application in nanoscale devices in the future technologies \cite{Liu2016}.

In the field of superconductivity the discovery of  the phenomenon of high-temperature superconductivity (HTSC) in iron-based materials has triggered a tremendous amount of innovative scientific efforts  \cite{Kamihara2008,Stewart2011,Johnson2015}. Iron chalcogenides are among VdW materials and are structurally the simplest high-temperature superconductors \cite{Hsu2008}. Owing to their simple crystal structure they have been considered as model systems for understanding the phenomenon of HTSC. The remarkable discovery of  HTSC in FeSe ML grown on SrTiO$_3$(001) [hereafter STO(001) or STO] with the highest transition temperature among all iron-based superconductors has made this system as one of the most attractive systems in condensed-matter physics \cite{Wang2012a,Liu2012,Tan2013,He2013,Bozovic2014,Ge2014,Lee2014}.
Although the physical mechanism leading to HTSC in FeSe ML on STO is not yet fully understood, it is generally believed that the superconductivity in this hybrid system is largely enhanced by interfacial effects \cite{Peng2014,Zhang2017,Zhao2018,Zhou2018,Song2019,Xu2020}.
It has been observed by means of several experimental techniques, including angle-resolved photoemission spectroscopy, that FeSe ML is strongly electron doped \cite{He2013,Seo2016,Zhang2017,Zhang2017a,Shi2017,Song2019}. The origin of this large carrier density inside FeSe ML has been postulated to be due to the charge transfer from the substrate into the film \cite{Zhou2016,Zhao2018}. However, a direct and unambiguous evidence of such a large charge transfer has not been reported experimentally.

Here by probing the dynamic charge response of the FeSe/STO interface, by means of high-resolution spectroscopy of slow electrons, we show that the experimentally measured frequency and momentum resolved dynamic charge response cannot be explained by assuming a simple film/substrate model only. Our detailed analysis unambiguously identifies a charge free depletion region in the STO substrate at the interface with FeSe ML. The existence  of such a thick depletion layer is accompanied with  a considerably large charge transfer from the STO into FeSe ML. The presence of the depletion region along with the large interfacial charge transfer leads to a substantial band bending and renormalization of the electronic bands at the interface. In addition to the fact that our findings contribute to the understanding of HTSC in FeSe ML, they would provide guidelines for designing new high-temperature superconductors through interface engineering. Moreover, the presence of a depletion layer at the interface is rather general and is expected to be observed in many combinations of VdW MLs put in contact with dielectric oxides or semiconducting substrates.

\section{Results and discussions}\label{Sec:resultsdiscuss}

The frequency and momentum resolved dynamic charge response of the epitaxial FeSe MLs grown on Nb-doped STO(001) (hereafter Nb-STO) was probed by means of high-resolution electron energy-loss spectroscopy (HREELS) (see Sec. \ref{Sec:Exp} of Materials and Methods for details), using slow electrons. Generally, in the electron scattering experiments from surfaces the scattering near the specular geometry is governed by the dipolar scattering mechanism \cite{Ibach1982}. In this region, known as dipolar lobe, the incoming electron interacts with the total charge density of the sample via  dipolar interactions. This means that the scattering intensity includes information from both electrons as well as the ions in the sample. The dipolar interaction is of Coulomb nature and hence is long range. Therefore, the scattered electrons carry information from the charge density fluctuations located not only near the surface region but also far below the surface, depending on their kinetic energy. The lower the kinetic energy the more surface sensitivity \cite{Ritz1984,Schaich1984,Lueth1988}. Since in our experiments we are mainly interested in the properties of FeSe ML and the interfacial region, we use electrons with kinetic energies as low as 4 to 8 eV.

\setcounter{figure}{0}
\makeatletter
\renewcommand{\figurename}{\textbf{Figure}}
\renewcommand{\thefigure}{\textbf{\@arabic\c@figure}}
\makeatother
 \begin{figure*}[t!]
	\centering
	\includegraphics[width=0.9\columnwidth]{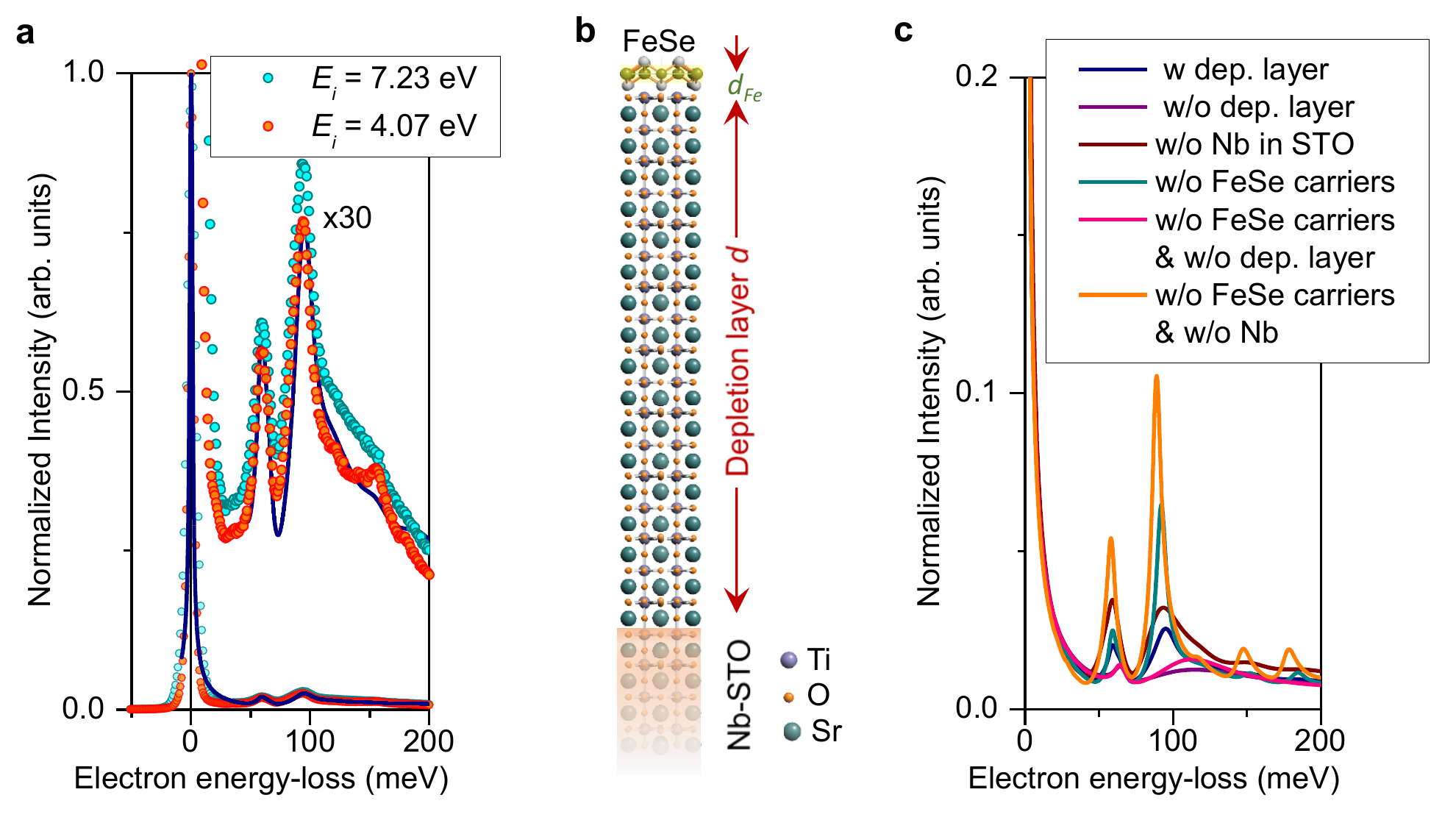}
	\caption{\textbf{Evidence of the depletion layer in high-resolution electron energy-loss spectra.}  \textbf{a} The solid circles represent the experimental spectra recorded at the specular geometry i.e., the wavevector of $q_{\parallel}=0$ \AA$^{-1}$, at the high symmetry $\bar{\Gamma}$--point and using two different incident beam energies ($E_i=4.07$ eV, orange and $E_i=7.23$ eV, light-blue colour). The simulated spectrum for $E_i=4.07$ eV is represented as the solid line. \textbf{b} The geometrical structure used for the simulation. In this model the Fe plane in FeSe ML is placed in a distance $d_{FeSe}=0.43$  nm above the STO(001) surface. The thickness of the depletion layer was varied to obtain the best fit to the experimental spectra ($d=6.5$ nm in this case). \textbf{c} Simulated spectra for different cases as denoted in the legend, dark-blue: considering all the required terms (the same as the one in \textbf{a}), violet: without considering a depletion layer, brown: without considering the charge carriers in STO, green: without a Drude term describing the free carriers in FeSe ML, light-red: without considering free carriers in FeSe ML and without considering a depletion layer, orange: without considering free carriers in FeSe ML as well as in STO.}
	\label{Fig:Spectra}
\end{figure*}

Figure \ref{Fig:Spectra}\textbf{a} shows the HREEL spectra recorded on the surface of FeSe ML grown on Nb-STO(001). The spectra were recorded at a temperature of $T=15$ K, below the superconducting transition temperature of the sample and at two different incident beam energies, namely $E_i=4.07$ eV and $E_i=7.23$ eV. The spectral function $S(q_{\parallel}, \omega)$ measured by HREELS directly reflects the dynamical response of the collective charge excitations in the system and is directly proportional to the imaginary part of the dynamic charge susceptibility $\mathfrak{Im}\chi(q, \omega)$ \cite{Vig2017,Husain2019}. Since the electrons are scattered by the total charge distribution of the sample, the scattering intensity must include information regarding collective ionic excitations i.e, phonons as well as collective electronic excitations i.e., plasmons. Moreover, any type of excitation representing a hybrid mode should also be excited within this mechanism. In the measured spectra presented in Fig. \ref{Fig:Spectra}\textbf{a} one observes several interesting features. Besides the elastic peak at the energy-loss of $\hbar\omega=0$  (zero-loss peak, ZLP) there are small peaks at $\hbar\omega=11.8$, 20.5, 24.8 and 36.7 meV. These are the phonon peaks of FeSe ML, which match perfectly to those probed on FeSe(001) single crystals \cite{Gnezdilov2013,Zakeri2017,Zakeri2018} and also those reported for the FeSe films on Nb-STO(001) of different thicknesses \cite{Zhang2018}. The most prominent features are the so-called Fuchs-Kliewer (FK) phonons of the STO substrate, which appear at the energies of $\hbar\omega=59.3$ and 94.5 meV.

 In principle, electrons can also excite multiple quanta (higher order harmonics) of FK phonons. Hence one should observe these excitations at the multiple frequencies of the principle excitations. The most interesting observation is that unlike the bare STO surfaces \cite{Conard1993}, in the case of FeSe/STO the higher order harmonics of the FK modes are strongly suppressed. Note that within the dipolar-scattering mechanism for a semi-infinite substrate the intensities of dipolar losses normalized to the elastic peak intensity scale in a $1/\sqrt{E_i}$ manner, where $E_i$ is the energy of the incident electron beam. This means that the lower the incident energy the larger the amplitude of FK modes. The intensities of the higher order harmonics of FK modes obey a Poisson distribution \cite{Ibach1970,Ibach1982}. Hence, in the spectra recorded using such low incident energies one shall clearly observe both the FK modes and their higher order harmonics, if they are present in the system. We will see that the strong suppression of the higher order harmonics of FK modes is due to the presence of the free charge carriers in FeSe ML and in the deeper layers of the Nb-STO substrate.

In order to shed light on the origin of the observed phenomenon the measured HREEL spectra were simulated. The simulation is based on the dipolar scattering theory (see Sec. \ref{Sec:Theo} of Materials and Methods for details). Our analysis indicates that considering ML FeSe on a semi-infinite doped STO(001) cannot explain the experimental spectra. The best model explaining the experimental spectra is considering a system composed of one ML of FeSe on $17$ unit cells of charge free insulating STO(001) on top of a semi-infinite Nb-STO(001). In this model the Fe plane in FeSe ML is placed in a distance $d_{FeSe}=0.43$  nm above the STO(001) surface \cite{Peng2020}. The structure is schematically sketched in Fig. \ref{Fig:Spectra}\textbf{b}. Only in this way both the peak position and amplitude of the excitations associated with the FK modes agree with those measured experimentally, as demonstrated in Fig. \ref{Fig:Spectra}\textbf{a}.  Similar to the experiment the higher harmonics of the principle FK modes are strongly suppressed due to the presence of the free carriers in FeSe ML and in the interior part of the substrate, below the depletion region.

Spectra simulated for several other configurations are presented in Fig. \ref{Fig:Spectra}\textbf{c} for a comparison. While the dark-blue colour represents the case shown in Fig. \ref{Fig:Spectra}\textbf{a} (assuming the structure shown in Fig. \ref{Fig:Spectra}\textbf{b}), the violet and light-red colours represent cases in which no depletion layer was considered. In these cases due to the presence of the charge carriers in Nb-STO, mainly due to the Nb doping, the FK peaks are heavily damped and are blue shifted. Neither the energy nor the peak height match the experimental spectra. Now, if one does not consider the contribution of the carriers in  Nb-STO, the FK peaks undergo a redshift and become almost identical in amplitude, as shown by the brown spectrum in Fig. \ref{Fig:Spectra}\textbf{c}. Likewise, the free carriers in FeSe ML are also essential to be considered, as demonstrated by the green spectrum.
In order to carefully investigate the influence of the depletion layer's thickness  $d$ on the spectra, simulations were performed for various values of $d$. Such data are presented in Supplementary Figure 1. It turned out that the best agreement with the experimental data can be achieved when $d$ is assumed to be $6.5\pm 1$ nm.

Another important result of the data presented in Fig. \ref{Fig:Spectra}\textbf{c} is that the higher order harmonics can only show up in the spectra when no charge carriers are present in the system. In such a case the FK peaks are rather sharp and intense and their higher order harmonics are clearly visible (see orange spectrum in Fig. \ref{Fig:Spectra}\textbf{c}). This observation   demonstrates the important role of the free carries in the suppression of the higher order harmonics of the FK modes. This was confirmed experimentally by performing experiments on a clean STO(001) surface at $T=180$ K. The experimental spectrum recorded over a wide range of energy-loss is presented in Supplementary Figure 2. The STO(001) surface was treated in exactly the same way as for the FeSe/STO samples, prior to the film deposition (see Sec. \ref{Sec:Exp} of Materials and Methods). In the absence of FeSe ML one observes the higher harmonics of the principle FK modes, indicating the role of FeSe ML's charge carriers in the suppression of these modes.  We note that the effective mass of bulk carriers inside STO is strongly temperature dependent and increases quickly with temperature \cite{Galzerani1982,Gervais1993,Conard1993,Collignon2020}. In order to reduce the effect of these bulk carriers on the spectra the experiments were performed at $T=180$ K, where the effective mass is supposed to be large.  The plasma frequency is directly proportional to  the square root of the carrier density and inversely proportional to the square root of the effective mass. Therefore, one expects a smaller plasma frequency associated with the bulk carriers in STO  at higher temperatures, assuming a temperature independent carrier concentration. This means that the effect of  temperature on the plasma frequency of the STO bulk carriers is equivalent to lowering the carrier concentration.

While performing HREELS experiments on FeSe films of various thicknesses it has been observed that the intensities of FK modes decrease exponentially with the thickness of the film \cite{Zhang2016}. This observation is interpreted as an indication of a large penetration depth of the dynamic electric fields associated with these modes into the FeSe films \cite{Zhang2016}. In order to investigate any possible screening of the dynamical electric field of the FK modes by  FeSe ML, spectra were recorded for various in-plane wavevectors $q_{\parallel}$. It was observed that the intensities of the FK modes decreases by three orders of magnitude while increasing $q_{\parallel}$ from zero to 0.6 $\AA^{-1}$ (from the $\bar{\Gamma}$-- towards the $\bar{\rm X}$--point of the surface Brillouin zone). At the same time the intensity of the ZLP also drops rapidly in a similar manner. Generally the profile of ZLP as a function of $q_{\parallel}$ is determined by the presence of defects at the surface. Since in the experiment the incident energy is low (only a few eV) the intensity profile of ZLP shall reflect the surface quality and the surface roughness. In the case of ultrathin films grown on a substrate, such as our samples, the roughness is almost entirely caused by the steps. The strong $q_{\parallel}$-dependent of the ZLP profile is the signature of low surface roughness and relatively wide terraces. We estimate an average width of at least 100 nm, in agreement with what was observed in our scanning tunneling microscopy studies \cite{Jandke2019}.
The decrease of the intensity of the FK modes with $q_{\parallel}$ is almost identical to that of ZLP. This is a strong indication that the observed FK modes at the off-specular geometry are excited via the same mechanism as those at the specular geometry, i.e., the dipolar scattering mechanism. In order to verify this hypothesis we performed simulations of the HREEL spectra for different geometries as in the experiment.

 \begin{figure*}[t!]
	\centering
	\includegraphics[width=0.5\columnwidth]{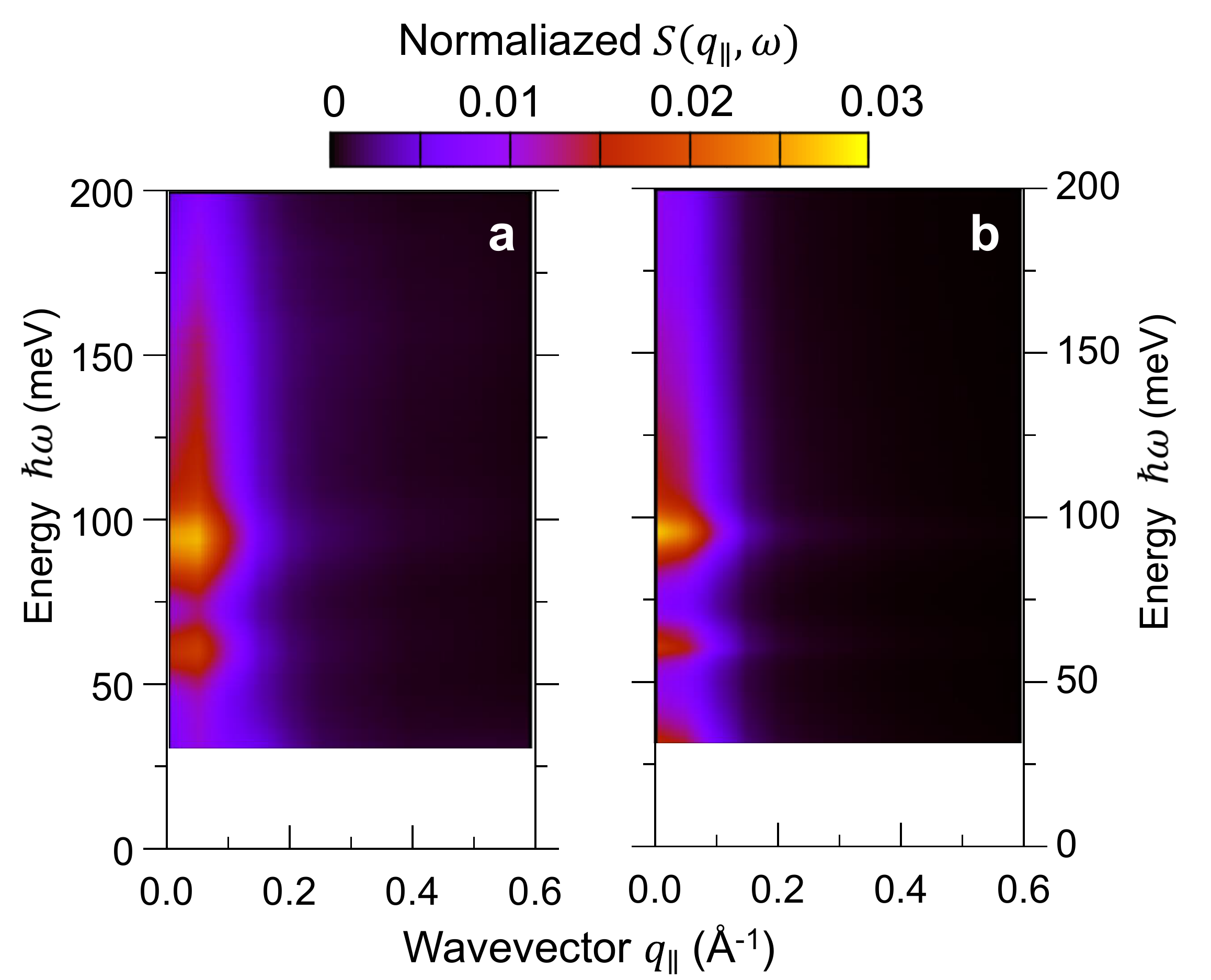}
	\caption{\textbf{The frequency and momentum dependent dynamic charge response.} \textbf{a} The experimental structural factor probed as a function of in-plane momentum and frequency on an FeSe ML grown on Nb-STO(001). This quantity is directly proportional to the imaginary part of dynamical charge susceptibility of the system. The data are recorded at a beam energy of $E_i=7.23$ eV. The simulated spectral function is shown in \textbf{b}. The values are normalized to the intensity of the zero-loss peak. The structure used for the simulation is the one shown in Fig. \ref{Fig:Spectra}\textbf{b}.}
	\label{Fig:DynamicRes}
\end{figure*}

In the experiment the spectra are recorded at the off-specular geometry and both the incident $\theta_i$ and scattered $\theta_f$ angles are adjusted to achieve the desired $q_{\parallel}$. Since the dipolar scattering theory is only valid in the vicinity of the specular geometry, the condition $\theta_i=\theta_f$ must be satisfied. Therefore, in the simulations only $\theta_i$ is adjusted to the experimental value. In order to account for the intensity drop of ZLP, the profile of the experimental ZLP is used in the simulation and the results are summarized in Fig. \ref{Fig:DynamicRes}. The experimental data are shown in Fig. \ref{Fig:DynamicRes}\textbf{a} while the results of simulations are presented in Fig. \ref{Fig:DynamicRes}\textbf{b}. The agreement between these two strongly suggests that the FK modes observed at off-specular angles are excited via the dipolar scattering mechanism, as in the specular geometry.

The dynamic electric fields generated by fluctuating electric dipoles associated with the FK modes are rather long range. Electrons scattered from the surface would feel these fields far above the surface. This is due to the long range nature of the Coulomb interaction. The angle under which the electrons are impinged onto the surface and consequently the parallel momentum of the incoming electron beam are rather unimportant. It seems that the screening effects of ML FeSe on dynamic electric fields is rather small. Such effects are essential to be considered also in other electron spectroscopy experiments e.g., angle-resolved photoemission spectroscopy (ARPES) \cite{Lee2014,Wang2016,Zhang2017,Zhou2017,Zhang2017,Faeth2021,Rademaker2021,Liu2021}. For instance, whether the observation of the so-called replica bands in the  ARPES experiments can be considered as a solid evidence for a strong electron-phonon coupling in this system requires further justifications \cite{Li2018}.

Looking at the data presented in Fig. \ref{Fig:DynamicRes} one observes a broad distribution of $S(q_{\parallel}, \omega)$  in frequency, in particular for $\hbar\omega >95$ meV. This is a consequence of  the free carriers in FeSe ML as well as in the inner part of Nb-STO. The presence of the charge carriers inside the Nb-STO substrate leads to a broadening and a blueshift of the FK modes (see for example the violet spectra in Fig. \ref{Fig:Spectra}\textbf{c}). In fact in such a situation the observed peaks are hybrid modes of both electronic and ionic collective excitations. The dynamic electric fields caused by the fluctuation of FK electrical dipoles are screened by the free carriers. One of the consequences of the depletion region at the interface is to enhance the effective dynamic electric field felt by the electrons scattered off (and above) the surface. These electric fields originate mainly from the insulating STO within the depletion region.

The observed depletion layer must be tightly connected to the charge transfer from the surface region of Nb-STO into FeSe ML. It has been observed by ARPES as well as tunneling spectroscopy that FeSe ML is heavily electron doped (0.12 $e^{-}$/Fe atom) \cite{He2013,Seo2016,Zhang2017,Zhang2017a,Shi2017,Song2019}. One plausible explanation for such a large charge density is that the carriers are transferred from the top several unit cells of Nb-STO into FeSe ML. Doping has been found to greatly enhance the superconducting transition temperature of both bulk as well as thick FeSe films \cite{Qian2011,Zhang2011,Miyata2015,Lei2016,Seo2016}. In order to verify that the charge transfer and the depletion layer are intimately interconnected and to see the consequences of the presence of the depletion layer on the electronic bands the system was modeled in a similar way suggested in Refs. \cite{Stengel2011,Reich2015,Zhou2016}. In this model the heavily doped Fe atomic plane in FeSe ML is considered as a charged sheet placed in the distance $d_{Fe}$ above the STO(001) (the structure shown in Fig. \ref{Fig:Spectra}\textbf{b}). The charged sheet generates an electric field and consequently a displacement vector $\vec{D}$, which is a function of distance from the charge sheet $z$ and extends into the depletion region.
After considering the boundary conditions the generalized displacement vector in the model system sketched in Fig. \ref{Fig:Spectra}\textbf{b} can be written as

\begin{equation}
\vec{D}(z)=\begin{cases}
\sigma \hat{z}=\mathcal{D} \hat{z} &\text{for }  0< z \leq d_{Fe},\\
\mathcal{D}\left(1-\frac{z}{\mathfrak{z}}\right) \hat{z} & \text{for } d_{Fe} \leq z < d,\\
  \end{cases}
\label{eq:displacement}
\end{equation}
where $\mathcal{D}=\sigma$ is the surface charge density of the Fe plane and is given by $\mathcal{D}=0.24e/a^2$ (here $a$ represents the in-plane lattice constant of the Fe plane which is the same as the one of  FeSe ML). $d$ is the thickness of the depletion layer after the charge transfer. $\mathfrak{z}$ is a constant in the units of $z$. It is given by the ratio of the surface charge density of the sheet and the volume charge density inside Nb-STO (see below). Note that in this model $z=0$ is placed on the Fe plane and $\hat{z}$ is pointing towards the inner part of Nb-STO(001).

At an infinitesimal distance $\delta$ just above the STO surface the displacement vector is given by $D(z=d_{Fe}-\delta)=\sigma$. On the other hand just below the surface inside STO at $z=d_{Fe}+\delta$, the displacement vector is given by $\vec{\nabla}\cdot\vec{D}(z)\mid_{z=d_{Fe}+\delta}=-en$, where $n$ is the carrier density inside Nb-STO. It includes both the charge carriers as a result of Nb doping as well as the defect induced carriers inside bulk Nb-STO, if there is any. The substrates used in our study are doped with 0.6\% Nb. In the simulated HREEL spectra the contribution of the charge carriers appears as a plasmon contribution to the effective dielectric function (see Sec. \ref{Sec:Theo} of Materials and Methods for a detailed description). The plasma frequency used for the simulation was 83 meV, showing the best agreement to the experimental spectra.  This value is also in agreement with the value reported earlier for Nb-STO using optical techniques \cite{Galzerani1982,Gervais1993}. Assuming that at low temperatures the effective mass of the carriers in Nb-STO is four times of the free electron's mass $m_{\rm eff}=4 m_e$ \cite{,Gervais1993,Collignon2020} one can estimate the carrier concentration $n$ in the interior part of Nb-STO, being about $n=1.18\times 10^{26}$ m$^{-3}$.
In order to estimate $\mathfrak{z}$ one may use the boundary conditions e.g., the field continuity principle at the interface. The field continuity at the interface [$D(z=d_{Fe}-\delta)=D(z=d_{Fe}+\delta)$] implies that $\mathfrak{z}=\sigma/n\simeq34$ unit cells. The thickness of the depletion layer after the charge transfer shall, however, be smaller than this value. The value found by analyzing the HREEL spectra was $d\simeq6.5 \pm 1$ nm ($\sim17 \pm 3$ unit cells).

 \begin{figure*}[t!]
	\centering
	\includegraphics[width=0.85\columnwidth]{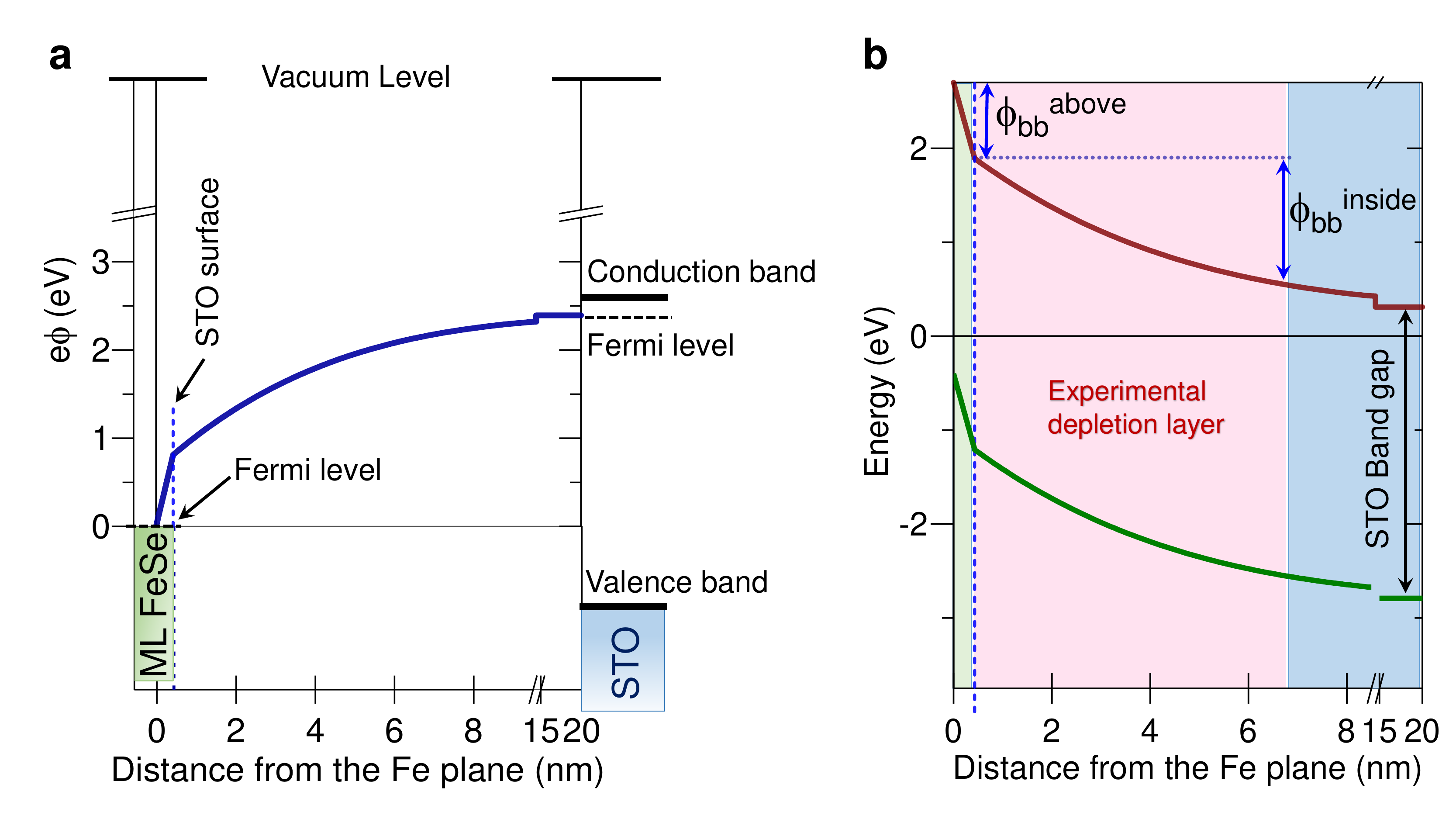}
	\caption{\textbf{Charge transfer and band bending at the interface between FeSe ML and STO.} \textbf{a} A schematic representation of the electronic bands of  FeSe ML and those of the Nd-STO when they are far apart. The electronic bands of  FeSe ML are filled up to the Fermi level, shown by the shaded green area.  In the case of STO the filled states are shown by the shaded blue area. The curve represents the calculated potential profile due to the charge transfer as a function of the distance from the Fe plane, when these two are brought in contact. The Fermi levels are shown by the dashed horizontal lines. The reference is put on the Fermi level of FeSe ML. \textbf{b} The band bending of the STO substrate in the vicinity of the interface, above $e\phi_{bb}^{\rm above}=0.8$ eV and inside $e\phi_{bb}^{\rm inside}=1.3$ eV the depletion layer. The valence and the conduction bands are shown by the green and brown colours. The depletion region is indicated by the shaded light-red colour and the inner part of the Nb-STO is shown by the shaded blue area.}
	\label{Fig:bandbending}
\end{figure*}

In order to estimate the band bending above the STO surface and inside the depletion layer the potential profile $\phi(z)$ should be calculated based on $\vec{D}(z)=-\varepsilon_{STO}[D(z)] \vec{\nabla} \phi(z)$. We note that due to the ferroelectric behavior of STO its dielectric constant $\varepsilon_{STO}$ is no longer a constant and depends on $D$. Using the function suggested in Refs. \cite{Stengel2011,Zhang2017a} for $\varepsilon(D)$ the potential was calculated and the results are summarized in Fig. \ref{Fig:bandbending}. The STO band gap was assumed to be 3.2 eV.

When FeSe ML and Nb-STO are far apart they possess different workfunctions and their Fermi levels are located at different energies (for an illustration see the sketches in Fig. \ref{Fig:bandbending}\textbf{a}).  Now if one assumes that upon attaching these two materials FeSe ML is negatively charged, a potential will be built up at the interface. The profile of this potential is presented in Fig. \ref{Fig:bandbending}\textbf{a}. The magnitude of this potential is large enough to balance the Fermi levels of the two materials and, at the same time, lead to a charge transfer. Such a built-in potential leads to a band banding in the STO in the vicinity of the interface. The profile shown in Fig. \ref{Fig:bandbending}\textbf{b} shows how the valence and conduction band just above the STO surface and also in the depletion layer near the interface are altered. Our analysis indicates a band bending of the conduction and valence band of about $e\phi_{bb}^{\rm inside}=1.3$ eV inside the depletion region. This value is estimated based on the build-in potential profile and represents the potential difference of the STO surface and the edge of the depletion region.
The presence of the depletion layer at the FeSe/STO interface would also lead to a redistribution of the charge in the FeSe layer in the vicinity of the interface. The strong variation of the potential in the vicinity of the interface just above the STO surface and below the Fe plane is as large as $e\phi_{bb}^{\rm above}=0.8 $ eV.

In the case of bulk FeSe it has been discussed that a description solely based
on an electron-phonon coupling as the pairing mechanism cannot explain the experimentally observed critical
temperature. The pairing mechanism has, therefore, been attributed to other bosonic excitations e.g., spin excitations. This has been confirmed by means of inelastic neutron scattering experiments \cite{Wang2015,Chen2019}. The observed resonance mode has been regarded as a signature of  a superconducting state with sign-changing order parameter, indicating an unconventional pairing mechanism via spin fluctuations\cite{Phelan2009,Ye2013,Mazin2015,Schrodi2020,Acharya2021}.

In the case of FeSe ML on STO the electronic structure is significantly simpler. The Fermi
surface consists of only two electron bands near the zone  boundary. The hole pockets are located  below the Fermi level by about 80 meV. Hence it is not straightforward to imagine an electronic coupling via the well-known ($\pi$--$\pi$) spin fluctuations \cite{Pelliciari2021}. However, unconventional pairing states may emerge from other states within the electronic bands \cite{Graser2009,Linscheid2016}. As a matter of fact the signature of the resonance mode has been experimentally observed by means of tunneling spectroscopy experiments, indicating an unconventional pairing mechanism via spin fluctuations in this system \cite{Jandke2019, Liu2019}.

On the other hand observation of the so-called replica bands in ARPES experiments has been considered as
an indication of a phonon-mediated superconductivity in this system.
The oxygen optical phonons  localized at
the TiO$_2$ interface have been suggested to be responsible for the pairing mechanism \cite{Lee2014,Wang2016,Tan2013,Zhang2017,Faeth2021,Rademaker2021,Liu2021}. This suggestion is  based on probing the quasiparticle band dispersions. However, no solid evidence has been reported by probing the phononic excitations of the system.


 Beside the possible mechanism discussed above, it has been suggested that a cooperative effect of several bosonic excitations may be responsible for the high transition temperature of this system \cite{Song2019,Schrodi2020a,Rademaker2021}.

Irrespective of the pairing mechanism responsible for the  superconductivity, the observed depletion layer and the large band renormalizations at the interface have very important consequences on the properties of FeSe ML \cite{Zhou2016,Zhou2017}.

\section{Conclusions}\label{Sec:conclutions}

By probing the dynamic charge response of the FeSe superconducting monolayer on STO(001), we identifed a charge depletion layer at the interface. The formation of the depletion layer explains the long-standing question regarding the origin of  the large charge density in FeSe ML, when it is grown on STO(001).  Moreover, the observed depletion layer and its consequences on the electronic properties of ML FeSe on STO(001) is a general phenomenon. We anticipate that such an effect exists also at the interface with many other oxide substrates, allowing an interfacial engineering of the superconducting states by growing ML FeSe on other dielectric surfaces or by combining MLs of other high-temperature superconducting materials with dielectric oxides. Moreover, the observed effect is also of great importance for nonsuperconducting VdW MLs put in contact with oxide or semiconducting substrates. In a similar way the exotic properties of these VdW MLs may be altered/tuned through interfacing with dielectric oxides.

\section{Methods}\label{Sec:Methods}
\subsection{Experiments}\label{Sec:Exp}

\subsubsection{Sample preparation and characterisation }
All the experiments were performed under ultrahigh vacuum conditions. Monolayers of FeSe were grown by means of molecular beam epitaxy technique on the Nb-STO(001) substrates \cite{Jandke2019,Li2014}. The Nb-doping level was 0.6\%. Prior to the film growth, the substrate was annealed at temperatures up to 1000 $^{\circ}$C and was then etched by selenium for 20 minutes. The sample was kept at the elevated temperature for 30 minutes and then was gradually cooled down to 480 $^{\circ}$C. Fe and Se were deposited simultaneously with a growth rate of 0.059ML/min at 480 $^{\circ}$C at a flux ratio of Fe:Se $\equiv$ 1:10. After deposition, the sample was annealed at 500 $^{\circ}$C for several hours to ensure a good morphological quality and desorption of the residual Se atoms. The growth was monitored by reflection high energy electron diffraction during the growth.

The sample was transferred using an ultrahigh vacuum suitcase to the scanning tunneling microscopy and high-resolution electron energy-loss spectroscopy chambers for further studies.

The morphological and electronic properties of the MLs were investigated by means of scanning tunneling microscopy \cite{Jandke2019}, showing a superconducting gap of $\Delta=11\pm3$ meV at 0.9 K.

\subsubsection{Probing the dynamic charge response}

The dynamic charge response of the system was probed by means of high-resolution electron energy-loss spectroscopy. The spectra were recorded along the main symmetry direction $\bar{\Gamma}$--$\bar{\rm X}$ of the surface Brillouin zone. The incident electron energy $E_i$ was between 4 and 8 eV. The incident energy was chosen such that the largest signal of FK modes is observed. The energy resolution  (half width at half maximum)  was between  1.5 and 4 meV. The wavevector $q_{\parallel}$ was varied by changing the scattering geometry i.e., by varying the angle between the incident and scattered beam with respect to the surface normal. The momentum resolution in the experiment was about $\Delta q=0.03$ \AA$^{-1}$ \cite{Ibach2003,Vollmer2003,Zakeri2014}.


\subsection{Simulation}\label{Sec:Theo}

In order to simulate the spectra a numerical scheme was developed based on the dipolar scattering theory \cite{Zakeri2021}. Within this formalism the single loss event is given by

\begin{equation}
P(\omega)= \frac{2}{a_0 k_i \cos\theta_i}\frac{1}{\omega}\mathfrak{Im}\left(\frac{-1}{\varepsilon_{\rm eff}(q, \omega)+1}\right),
\label{Eq:Singleloss}
\end{equation}
where $k_i=\sqrt{2m_eE_i}/\hbar$ is the momentum of the incident electron beam impinging onto the surface under the incident angle $\theta_i$. $m_e$ is the electron mass and $\hbar$ is the reduced Planck's constant. $a_0$ is the Bohr radius and $\varepsilon_{\rm eff}(q, \omega)$ represents the effective dielectric function of the medium.  For layered systems, such as our case, $\varepsilon_{\rm eff}(q, \omega)$ may be written in terms of the dielectric functions of individual layers $\varepsilon^{p}(q,\omega)$, where $p$ is the layer index. For that a multi-slab representation can be used and $\varepsilon_{\rm eff}(q, \omega)$ can be expanded in series containing $\varepsilon^{p}(q,\omega)$ (see for example Refs.  \cite{Sunjic1971,Lucas1972,Lambin1990,Lazzari2018} for details). It is important to note that Eq. (\ref{Eq:Singleloss}) describes only the single-loss probability for an electron having a wavevector $k_\mathrm{i}$ to be scattered from a semi-infinite slab system and lose the energy $\omega$ at $T=0$ K. The multiple scattering events, the elastic peak, and temperature effects were included using the approach originally developed by Lucas and \v{S}unji\'{c} \cite{Sunjic1971,Lucas1972,Lambin1990,Lazzari2018}.

In the constructed multi-slab model the dielectric function of each individual layer may be written in terms the different contributions
\begin{equation}
\varepsilon^{p}(q,\omega) = \varepsilon^{p}_{\infty} + \varepsilon^{p}_\mathrm{phonon} + \varepsilon^{p}_\mathrm{plasmon}.
\label{eq:dielectric_function}
\end{equation}

The first term is the background dielectric constant of the layer and is independent of momentum and frequency. We use the values of 15 and 5.9 for FeSe and STO, respectively  \cite{Yuan2012,Gervais1993,Zhou2017,Zhou2016}. The second term represents the lattice contribution i.e., the phonons and the third term represents the contribution of the free carriers i.e., plasmons to the total dielectric constant of each individual layer.

The phononic contribution to the dielectric constant of each layer is given by

\begin{equation}
\varepsilon^{p}_\mathrm{phonon} = \sum\limits_{j=1}^{j=m} \frac{Q_j \omega_{\mathrm{TO},j}^2}{\omega_{\mathrm{TO},j}^2 -\omega^2 - \mathrm{i} \gamma_{\mathrm{TO},j} \omega}.
\label{eq:ph-oscillators-sum}
\end{equation}

Here $m$ is the number of all transversal optical (TO) modes. Each phonon mode is assumed to act as an oscillator with a dimensionless oscillator strength $Q_j$, which, in turn, depends on the splitting between TO and longitudinal optical (LO) modes

\begin{equation}
Q_j = \frac{\varepsilon_\infty}{\omega_{\mathrm{TO},j}^2} \frac{\prod\limits_{l} \left(\omega_{\mathrm{LO},l}^2 - \omega_{\mathrm{TO},j}^2 \right)}{\prod\limits_{l\neq j}\left(\omega_{\mathrm{TO},l}^2 - \omega_{\mathrm{TO},j}^2 \right)},
\label{eq:oscillators-strength}
\end{equation}
where $\omega_{\mathrm{TO},j}$ and $\omega_{\mathrm{LO},j}$ are the frequency of the $j$-th TO and LO phonon modes, respectively. $\gamma_{\mathrm{TO},j}$ and $\gamma_{\mathrm{LO},j}$ represent their corresponding damping.

In order to account for the charge carries a Drude-like term was considered in the dielectric function of each layer

\begin{equation}
\varepsilon^{p}_\mathrm{plasmon} = -\varepsilon_\infty\frac{ \omega_{pl}^2-i(\gamma_{pl}-\gamma_0)\omega}{\omega(\omega+i \gamma_0)} \, ,
\label{eq:Drude}
\end{equation}
where $\omega_{pl}$ is the plasma frequency associated with the charge carriers and is directly related to the carrier density $n_c$, the effective mass $m_{\rm eff}$ and the vacuum permittivity $\epsilon_0$ by $\omega_{pl}^2 = \frac{n_c e^2}{\varepsilon_\infty \epsilon_0 m_{\rm eff}}$. The quantities $\gamma_{pl}$ and $\gamma_0$ describe the linewidth of the plasmon peak and are determined by the plasmon relaxation time. In case that $\gamma_{pl} = \gamma_0$, Eq. (\ref{eq:Drude}) will be simplified to the well-known Drude term.

For the Nb-STO we use the form introduced in Eq. (\ref{eq:Drude}) with $\omega_{pl}=83$ meV, $\gamma_{pl} =75$ meV and $\gamma_0=5$ meV. The values are estimated by extrapolating the values measured by optical techniques at liquid nitrogen temperature to our measurement temperature (T=15 K) \cite{Gervais1993,Eagles1996}. The extrapolation is based on the temperature dependence of effective mass as discussed in detail in Ref. \cite{Collignon2020}. Using these values the best agreement with the experimental spectra was achieved. In the case of FeSe ML we use a Drude form ($\gamma_{pl} = \gamma_0$). The value used for the plasma frequency was $\omega_{pl}=334$ meV estimated based on $\omega_{pl}^2 = \frac{n_c e^2}{\varepsilon_\infty \epsilon_0 m_{\rm eff}}$, assuming $m_{\rm eff}\simeq3$ and $n_c=0.12$ $e^{-}$/Fe .

The phonon and plasmon frequencies and their damping serve as the input of the simulations. The values used for your simulations are tabulated in Supplementary Table 1.

\section*{Data availability}
The datasets generated and/or analysed during the current study are available from the corresponding author on reasonable request.

\section*{Code availability}
The codes associated with this manuscript are available from the corresponding
author on reasonable request.

\section*{Author contributions}
Kh.Z. initiated the idea of the study, supervised the whole project, conceived and planned the experiments, analyzed the experimental data, participated in the development of the simulations code, developed the model for calculation of the potential profile, performed the calculations and simulations and wrote the paper. D.R. contributed to carrying out the HREELS experiments. M.D. and J.W. developed the simulation code. D.R., F.Y., J.J. and W.W. prepared the samples and performed the STM experiments. J.S. commented on the models used in the manuscript.

\section*{Competing interests}
The authors declare no competing interests.

\section*{Additional information}
Correspondence and requests for materials should be addressed to Kh.Z (khalil.zakeri@kit.edu).

\section*{Acknowledgements}
Kh.Z. acknowledges funding from the Deutsche Forschungsgemeinschaft (DFG) through the Heisenberg Programme ZA 902/3-1 and ZA 902/6-1 and the DFG Grant No. ZA 902/5-1.  The research of J.J. and W.W. was supported by DFG through Grant No.
Wu 394/12-1. The research of J.S. was supported by DFG via Grant No. SCHM 1031/7-1.
F.Y.  acknowledges funding from the Alexander von Humboldt Foundation. Kh.Z. thanks the Physikalisches Institut for hosting the group and providing the necessary infrastructure.

\bibliography{Refs}

\begin{thebibliography}{78}%
\makeatletter
\providecommand \@ifxundefined [1]{%
 \@ifx{#1\undefined}
}%
\providecommand \@ifnum [1]{%
 \ifnum #1\expandafter \@firstoftwo
 \else \expandafter \@secondoftwo
 \fi
}%
\providecommand \@ifx [1]{%
 \ifx #1\expandafter \@firstoftwo
 \else \expandafter \@secondoftwo
 \fi
}%
\providecommand \natexlab [1]{#1}%
\providecommand \enquote  [1]{``#1''}%
\providecommand \bibnamefont  [1]{#1}%
\providecommand \bibfnamefont [1]{#1}%
\providecommand \citenamefont [1]{#1}%
\providecommand \href@noop [0]{\@secondoftwo}%
\providecommand \href [0]{\begingroup \@sanitize@url \@href}%
\providecommand \@href[1]{\@@startlink{#1}\@@href}%
\providecommand \@@href[1]{\endgroup#1\@@endlink}%
\providecommand \@sanitize@url [0]{\catcode `\\12\catcode `\$12\catcode
  `\&12\catcode `\#12\catcode `\^12\catcode `\_12\catcode `\%12\relax}%
\providecommand \@@startlink[1]{}%
\providecommand \@@endlink[0]{}%
\providecommand \url  [0]{\begingroup\@sanitize@url \@url }%
\providecommand \@url [1]{\endgroup\@href {#1}{\urlprefix }}%
\providecommand \urlprefix  [0]{URL }%
\providecommand \Eprint [0]{\href }%
\providecommand \doibase [0]{https://doi.org/}%
\providecommand \selectlanguage [0]{\@gobble}%
\providecommand \bibinfo  [0]{\@secondoftwo}%
\providecommand \bibfield  [0]{\@secondoftwo}%
\providecommand \translation [1]{[#1]}%
\providecommand \BibitemOpen [0]{}%
\providecommand \bibitemStop [0]{}%
\providecommand \bibitemNoStop [0]{.\EOS\space}%
\providecommand \EOS [0]{\spacefactor3000\relax}%
\providecommand \BibitemShut  [1]{\csname bibitem#1\endcsname}%
\let\auto@bib@innerbib\@empty
\bibitem [{\citenamefont {Geim}\ and\ \citenamefont
  {Grigorieva}(2013)}]{Geim2013}%
  \BibitemOpen
  \bibfield  {author} {\bibinfo {author} {\bibfnamefont {A.~K.}\ \bibnamefont
  {Geim}}\ and\ \bibinfo {author} {\bibfnamefont {I.~V.}\ \bibnamefont
  {Grigorieva}},\ }\bibfield  {title} {\bibinfo {title} {Van der waals
  heterostructures},\ }\href {https://doi.org/10.1038/nature12385} {\bibfield
  {journal} {\bibinfo  {journal} {Nature}\ }\textbf {\bibinfo {volume} {499}},\
  \bibinfo {pages} {419} (\bibinfo {year} {2013})}\BibitemShut {NoStop}%
\bibitem [{\citenamefont {Novoselov}\ \emph {et~al.}(2016)\citenamefont
  {Novoselov}, \citenamefont {Mishchenko}, \citenamefont {Carvalho},\ and\
  \citenamefont {Neto}}]{Novoselov2016}%
  \BibitemOpen
  \bibfield  {author} {\bibinfo {author} {\bibfnamefont {K.~S.}\ \bibnamefont
  {Novoselov}}, \bibinfo {author} {\bibfnamefont {A.}~\bibnamefont
  {Mishchenko}}, \bibinfo {author} {\bibfnamefont {A.}~\bibnamefont
  {Carvalho}},\ and\ \bibinfo {author} {\bibfnamefont {A.~H.~C.}\ \bibnamefont
  {Neto}},\ }\bibfield  {title} {\bibinfo {title} {{2D} materials and van der
  waals heterostructures},\ }\href {https://doi.org/10.1126/science.aac9439}
  {\bibfield  {journal} {\bibinfo  {journal} {Science}\ }\textbf {\bibinfo
  {volume} {353}},\ \bibinfo {pages} {461} (\bibinfo {year}
  {2016})}\BibitemShut {NoStop}%
\bibitem [{\citenamefont {Jiang}\ \emph {et~al.}(2021)\citenamefont {Jiang},
  \citenamefont {Liu}, \citenamefont {Xing}, \citenamefont {Liu}, \citenamefont
  {Guo}, \citenamefont {Liu},\ and\ \citenamefont {Zhao}}]{Jiang2021}%
  \BibitemOpen
  \bibfield  {author} {\bibinfo {author} {\bibfnamefont {X.}~\bibnamefont
  {Jiang}}, \bibinfo {author} {\bibfnamefont {Q.}~\bibnamefont {Liu}}, \bibinfo
  {author} {\bibfnamefont {J.}~\bibnamefont {Xing}}, \bibinfo {author}
  {\bibfnamefont {N.}~\bibnamefont {Liu}}, \bibinfo {author} {\bibfnamefont
  {Y.}~\bibnamefont {Guo}}, \bibinfo {author} {\bibfnamefont {Z.}~\bibnamefont
  {Liu}},\ and\ \bibinfo {author} {\bibfnamefont {J.}~\bibnamefont {Zhao}},\
  }\bibfield  {title} {\bibinfo {title} {Recent progress on 2d magnets:
  Fundamental mechanism, structural design and modification},\ }\href
  {https://doi.org/10.1063/5.0039979} {\bibfield  {journal} {\bibinfo
  {journal} {Applied Physics Reviews}\ }\textbf {\bibinfo {volume} {8}},\
  \bibinfo {pages} {031305} (\bibinfo {year} {2021})}\BibitemShut {NoStop}%
\bibitem [{\citenamefont {Liu}\ \emph {et~al.}(2016)\citenamefont {Liu},
  \citenamefont {Weiss}, \citenamefont {Duan}, \citenamefont {Cheng},
  \citenamefont {Huang},\ and\ \citenamefont {Duan}}]{Liu2016}%
  \BibitemOpen
  \bibfield  {author} {\bibinfo {author} {\bibfnamefont {Y.}~\bibnamefont
  {Liu}}, \bibinfo {author} {\bibfnamefont {N.~O.}\ \bibnamefont {Weiss}},
  \bibinfo {author} {\bibfnamefont {X.}~\bibnamefont {Duan}}, \bibinfo {author}
  {\bibfnamefont {H.-C.}\ \bibnamefont {Cheng}}, \bibinfo {author}
  {\bibfnamefont {Y.}~\bibnamefont {Huang}},\ and\ \bibinfo {author}
  {\bibfnamefont {X.}~\bibnamefont {Duan}},\ }\bibfield  {title} {\bibinfo
  {title} {Van der waals heterostructures and devices},\ }\href
  {https://doi.org/10.1038/natrevmats.2016.42} {\bibfield  {journal} {\bibinfo
  {journal} {Nature Reviews Materials}\ }\textbf {\bibinfo {volume} {1}},\
  \bibinfo {pages} {16042} (\bibinfo {year} {2016})}\BibitemShut {NoStop}%
\bibitem [{\citenamefont {Kamihara}\ \emph {et~al.}(2008)\citenamefont
  {Kamihara}, \citenamefont {Watanabe}, \citenamefont {Hirano},\ and\
  \citenamefont {Hosono}}]{Kamihara2008}%
  \BibitemOpen
  \bibfield  {author} {\bibinfo {author} {\bibfnamefont {Y.}~\bibnamefont
  {Kamihara}}, \bibinfo {author} {\bibfnamefont {T.}~\bibnamefont {Watanabe}},
  \bibinfo {author} {\bibfnamefont {M.}~\bibnamefont {Hirano}},\ and\ \bibinfo
  {author} {\bibfnamefont {H.}~\bibnamefont {Hosono}},\ }\bibfield  {title}
  {\bibinfo {title} {Iron-based layered superconductor
  {La[O$_{1-x}$F$_{x}$]FeAs} (x = 0.0-0.12) with {T$_c$} = 26 {K}},\ }\href
  {https://doi.org/10.1021/ja800073m} {\bibfield  {journal} {\bibinfo
  {journal} {J. Am. Chem. Soc.}\ }\textbf {\bibinfo {volume} {130}},\ \bibinfo
  {pages} {3296} (\bibinfo {year} {2008})}\BibitemShut {NoStop}%
\bibitem [{\citenamefont {Stewart}(2011)}]{Stewart2011}%
  \BibitemOpen
  \bibfield  {author} {\bibinfo {author} {\bibfnamefont {G.~R.}\ \bibnamefont
  {Stewart}},\ }\bibfield  {title} {\bibinfo {title} {Superconductivity in iron
  compounds},\ }\href {https://doi.org/10.1103/RevModPhys.83.1589} {\bibfield
  {journal} {\bibinfo  {journal} {Rev. Mod. Phys.}\ }\textbf {\bibinfo {volume}
  {83}},\ \bibinfo {pages} {1589} (\bibinfo {year} {2011})},\ \bibinfo {note}
  {and references therein}\BibitemShut {NoStop}%
\bibitem [{\citenamefont {Johnson}\ \emph {et~al.}(2015)\citenamefont
  {Johnson}, \citenamefont {Xu},\ and\ \citenamefont {Yin}}]{Johnson2015}%
  \BibitemOpen
  \bibfield  {author} {\bibinfo {author} {\bibfnamefont {P.}~\bibnamefont
  {Johnson}}, \bibinfo {author} {\bibfnamefont {G.}~\bibnamefont {Xu}},\ and\
  \bibinfo {author} {\bibfnamefont {W.}~\bibnamefont {Yin}},\ }\href@noop {}
  {\emph {\bibinfo {title} {Iron-Based Superconductivity}}},\ Springer Series
  in Materials Science\ (\bibinfo  {publisher} {Springer International
  Publishing},\ \bibinfo {year} {2015})\BibitemShut {NoStop}%
\bibitem [{\citenamefont {Hsu}\ \emph {et~al.}(2008)\citenamefont {Hsu},
  \citenamefont {Luo}, \citenamefont {Yeh}, \citenamefont {Chen}, \citenamefont
  {Huang}, \citenamefont {Wu}, \citenamefont {Lee}, \citenamefont {Huang},
  \citenamefont {Chu}, \citenamefont {Yan} \emph {et~al.}}]{Hsu2008}%
  \BibitemOpen
  \bibfield  {author} {\bibinfo {author} {\bibfnamefont {F.-C.}\ \bibnamefont
  {Hsu}}, \bibinfo {author} {\bibfnamefont {J.-Y.}\ \bibnamefont {Luo}},
  \bibinfo {author} {\bibfnamefont {K.-W.}\ \bibnamefont {Yeh}}, \bibinfo
  {author} {\bibfnamefont {T.-K.}\ \bibnamefont {Chen}}, \bibinfo {author}
  {\bibfnamefont {T.-W.}\ \bibnamefont {Huang}}, \bibinfo {author}
  {\bibfnamefont {P.~M.}\ \bibnamefont {Wu}}, \bibinfo {author} {\bibfnamefont
  {Y.-C.}\ \bibnamefont {Lee}}, \bibinfo {author} {\bibfnamefont {Y.-L.}\
  \bibnamefont {Huang}}, \bibinfo {author} {\bibfnamefont {Y.-Y.}\ \bibnamefont
  {Chu}}, \bibinfo {author} {\bibfnamefont {D.-C.}\ \bibnamefont {Yan}}, \emph
  {et~al.},\ }\bibfield  {title} {\bibinfo {title} {Superconductivity in the
  pbo-type structure $\alpha$-fese},\ }\href
  {https://doi.org/10.1073/pnas.0807325105} {\bibfield  {journal} {\bibinfo
  {journal} {Proceedings of the National Academy of Sciences}\ }\textbf
  {\bibinfo {volume} {105}},\ \bibinfo {pages} {14262} (\bibinfo {year}
  {2008})}\BibitemShut {NoStop}%
\bibitem [{\citenamefont {Wang}\ \emph {et~al.}(2012)\citenamefont {Wang},
  \citenamefont {Li}, \citenamefont {Zhang}, \citenamefont {Zhang},
  \citenamefont {Zhang}, \citenamefont {Li}, \citenamefont {Ding},
  \citenamefont {Ou}, \citenamefont {Deng}, \citenamefont {Chang},
  \citenamefont {Wen}, \citenamefont {Song}, \citenamefont {He}, \citenamefont
  {Jia}, \citenamefont {Ji}, \citenamefont {Wang}, \citenamefont {Wang},
  \citenamefont {Chen}, \citenamefont {Ma},\ and\ \citenamefont
  {Xue}}]{Wang2012a}%
  \BibitemOpen
  \bibfield  {author} {\bibinfo {author} {\bibfnamefont {Q.-Y.}\ \bibnamefont
  {Wang}}, \bibinfo {author} {\bibfnamefont {Z.}~\bibnamefont {Li}}, \bibinfo
  {author} {\bibfnamefont {W.-H.}\ \bibnamefont {Zhang}}, \bibinfo {author}
  {\bibfnamefont {Z.-C.}\ \bibnamefont {Zhang}}, \bibinfo {author}
  {\bibfnamefont {J.-S.}\ \bibnamefont {Zhang}}, \bibinfo {author}
  {\bibfnamefont {W.}~\bibnamefont {Li}}, \bibinfo {author} {\bibfnamefont
  {H.}~\bibnamefont {Ding}}, \bibinfo {author} {\bibfnamefont {Y.-B.}\
  \bibnamefont {Ou}}, \bibinfo {author} {\bibfnamefont {P.}~\bibnamefont
  {Deng}}, \bibinfo {author} {\bibfnamefont {K.}~\bibnamefont {Chang}},
  \bibinfo {author} {\bibfnamefont {J.}~\bibnamefont {Wen}}, \bibinfo {author}
  {\bibfnamefont {C.-L.}\ \bibnamefont {Song}}, \bibinfo {author}
  {\bibfnamefont {K.}~\bibnamefont {He}}, \bibinfo {author} {\bibfnamefont
  {J.-F.}\ \bibnamefont {Jia}}, \bibinfo {author} {\bibfnamefont {S.-H.}\
  \bibnamefont {Ji}}, \bibinfo {author} {\bibfnamefont {Y.-Y.}\ \bibnamefont
  {Wang}}, \bibinfo {author} {\bibfnamefont {L.-L.}\ \bibnamefont {Wang}},
  \bibinfo {author} {\bibfnamefont {X.}~\bibnamefont {Chen}}, \bibinfo {author}
  {\bibfnamefont {X.-C.}\ \bibnamefont {Ma}},\ and\ \bibinfo {author}
  {\bibfnamefont {Q.-K.}\ \bibnamefont {Xue}},\ }\bibfield  {title} {\bibinfo
  {title} {Interface-induced high-temperature superconductivity in single
  unit-cell {FeSe} films on {SrTiO$_3$}},\ }\href
  {https://doi.org/10.1088/0256-307x/29/3/037402} {\bibfield  {journal}
  {\bibinfo  {journal} {Chinese Physics Letters}\ }\textbf {\bibinfo {volume}
  {29}},\ \bibinfo {pages} {037402} (\bibinfo {year} {2012})}\BibitemShut
  {NoStop}%
\bibitem [{\citenamefont {Liu}\ \emph {et~al.}(2012)\citenamefont {Liu},
  \citenamefont {Zhang}, \citenamefont {Mou}, \citenamefont {He}, \citenamefont
  {Ou}, \citenamefont {Wang}, \citenamefont {Li}, \citenamefont {Wang},
  \citenamefont {Zhao}, \citenamefont {He}, \citenamefont {Peng}, \citenamefont
  {Liu}, \citenamefont {Chen}, \citenamefont {Yu}, \citenamefont {Liu},
  \citenamefont {Dong}, \citenamefont {Zhang}, \citenamefont {Chen},
  \citenamefont {Xu}, \citenamefont {Hu}, \citenamefont {Chen}, \citenamefont
  {Ma}, \citenamefont {Xue},\ and\ \citenamefont {Zhou}}]{Liu2012}%
  \BibitemOpen
  \bibfield  {author} {\bibinfo {author} {\bibfnamefont {D.}~\bibnamefont
  {Liu}}, \bibinfo {author} {\bibfnamefont {W.}~\bibnamefont {Zhang}}, \bibinfo
  {author} {\bibfnamefont {D.}~\bibnamefont {Mou}}, \bibinfo {author}
  {\bibfnamefont {J.}~\bibnamefont {He}}, \bibinfo {author} {\bibfnamefont
  {Y.-B.}\ \bibnamefont {Ou}}, \bibinfo {author} {\bibfnamefont {Q.-Y.}\
  \bibnamefont {Wang}}, \bibinfo {author} {\bibfnamefont {Z.}~\bibnamefont
  {Li}}, \bibinfo {author} {\bibfnamefont {L.}~\bibnamefont {Wang}}, \bibinfo
  {author} {\bibfnamefont {L.}~\bibnamefont {Zhao}}, \bibinfo {author}
  {\bibfnamefont {S.}~\bibnamefont {He}}, \bibinfo {author} {\bibfnamefont
  {Y.}~\bibnamefont {Peng}}, \bibinfo {author} {\bibfnamefont {X.}~\bibnamefont
  {Liu}}, \bibinfo {author} {\bibfnamefont {C.}~\bibnamefont {Chen}}, \bibinfo
  {author} {\bibfnamefont {L.}~\bibnamefont {Yu}}, \bibinfo {author}
  {\bibfnamefont {G.}~\bibnamefont {Liu}}, \bibinfo {author} {\bibfnamefont
  {X.}~\bibnamefont {Dong}}, \bibinfo {author} {\bibfnamefont {J.}~\bibnamefont
  {Zhang}}, \bibinfo {author} {\bibfnamefont {C.}~\bibnamefont {Chen}},
  \bibinfo {author} {\bibfnamefont {Z.}~\bibnamefont {Xu}}, \bibinfo {author}
  {\bibfnamefont {J.}~\bibnamefont {Hu}}, \bibinfo {author} {\bibfnamefont
  {X.}~\bibnamefont {Chen}}, \bibinfo {author} {\bibfnamefont {X.}~\bibnamefont
  {Ma}}, \bibinfo {author} {\bibfnamefont {Q.}~\bibnamefont {Xue}},\ and\
  \bibinfo {author} {\bibfnamefont {X.}~\bibnamefont {Zhou}},\ }\bibfield
  {title} {\bibinfo {title} {Electronic origin of high-temperature
  superconductivity in single-layer {FeSe} superconductor},\ }\bibfield
  {journal} {\bibinfo  {journal} {Nature Communications}\ }\textbf {\bibinfo
  {volume} {3}},\ \href {https://doi.org/10.1038/ncomms1946}
  {10.1038/ncomms1946} (\bibinfo {year} {2012})\BibitemShut {NoStop}%
\bibitem [{\citenamefont {Tan}\ \emph {et~al.}(2013)\citenamefont {Tan},
  \citenamefont {Zhang}, \citenamefont {Xia}, \citenamefont {Ye}, \citenamefont
  {Chen}, \citenamefont {Xie}, \citenamefont {Peng}, \citenamefont {Xu},
  \citenamefont {Fan}, \citenamefont {Xu}, \citenamefont {Jiang}, \citenamefont
  {Zhang}, \citenamefont {Lai}, \citenamefont {Xiang}, \citenamefont {Hu},
  \citenamefont {Xie},\ and\ \citenamefont {Feng}}]{Tan2013}%
  \BibitemOpen
  \bibfield  {author} {\bibinfo {author} {\bibfnamefont {S.}~\bibnamefont
  {Tan}}, \bibinfo {author} {\bibfnamefont {Y.}~\bibnamefont {Zhang}}, \bibinfo
  {author} {\bibfnamefont {M.}~\bibnamefont {Xia}}, \bibinfo {author}
  {\bibfnamefont {Z.}~\bibnamefont {Ye}}, \bibinfo {author} {\bibfnamefont
  {F.}~\bibnamefont {Chen}}, \bibinfo {author} {\bibfnamefont {X.}~\bibnamefont
  {Xie}}, \bibinfo {author} {\bibfnamefont {R.}~\bibnamefont {Peng}}, \bibinfo
  {author} {\bibfnamefont {D.}~\bibnamefont {Xu}}, \bibinfo {author}
  {\bibfnamefont {Q.}~\bibnamefont {Fan}}, \bibinfo {author} {\bibfnamefont
  {H.}~\bibnamefont {Xu}}, \bibinfo {author} {\bibfnamefont {J.}~\bibnamefont
  {Jiang}}, \bibinfo {author} {\bibfnamefont {T.}~\bibnamefont {Zhang}},
  \bibinfo {author} {\bibfnamefont {X.}~\bibnamefont {Lai}}, \bibinfo {author}
  {\bibfnamefont {T.}~\bibnamefont {Xiang}}, \bibinfo {author} {\bibfnamefont
  {J.}~\bibnamefont {Hu}}, \bibinfo {author} {\bibfnamefont {B.}~\bibnamefont
  {Xie}},\ and\ \bibinfo {author} {\bibfnamefont {D.}~\bibnamefont {Feng}},\
  }\bibfield  {title} {\bibinfo {title} {Interface-induced superconductivity
  and strain-dependent spin density waves in {FeSe}/{SrTiO}3~thin films},\
  }\href {https://doi.org/10.1038/nmat3654} {\bibfield  {journal} {\bibinfo
  {journal} {Nature Materials}\ }\textbf {\bibinfo {volume} {12}},\ \bibinfo
  {pages} {634} (\bibinfo {year} {2013})}\BibitemShut {NoStop}%
\bibitem [{\citenamefont {He}\ \emph {et~al.}(2013)\citenamefont {He},
  \citenamefont {He}, \citenamefont {Zhang}, \citenamefont {Zhao},
  \citenamefont {Liu}, \citenamefont {Liu}, \citenamefont {Mou}, \citenamefont
  {Ou}, \citenamefont {Wang}, \citenamefont {Li}, \citenamefont {Wang},
  \citenamefont {Peng}, \citenamefont {Liu}, \citenamefont {Chen},
  \citenamefont {Yu}, \citenamefont {Liu}, \citenamefont {Dong}, \citenamefont
  {Zhang}, \citenamefont {Chen}, \citenamefont {Xu}, \citenamefont {Chen},
  \citenamefont {Ma}, \citenamefont {Xue},\ and\ \citenamefont
  {Zhou}}]{He2013}%
  \BibitemOpen
  \bibfield  {author} {\bibinfo {author} {\bibfnamefont {S.}~\bibnamefont
  {He}}, \bibinfo {author} {\bibfnamefont {J.}~\bibnamefont {He}}, \bibinfo
  {author} {\bibfnamefont {W.}~\bibnamefont {Zhang}}, \bibinfo {author}
  {\bibfnamefont {L.}~\bibnamefont {Zhao}}, \bibinfo {author} {\bibfnamefont
  {D.}~\bibnamefont {Liu}}, \bibinfo {author} {\bibfnamefont {X.}~\bibnamefont
  {Liu}}, \bibinfo {author} {\bibfnamefont {D.}~\bibnamefont {Mou}}, \bibinfo
  {author} {\bibfnamefont {Y.-B.}\ \bibnamefont {Ou}}, \bibinfo {author}
  {\bibfnamefont {Q.-Y.}\ \bibnamefont {Wang}}, \bibinfo {author}
  {\bibfnamefont {Z.}~\bibnamefont {Li}}, \bibinfo {author} {\bibfnamefont
  {L.}~\bibnamefont {Wang}}, \bibinfo {author} {\bibfnamefont {Y.}~\bibnamefont
  {Peng}}, \bibinfo {author} {\bibfnamefont {Y.}~\bibnamefont {Liu}}, \bibinfo
  {author} {\bibfnamefont {C.}~\bibnamefont {Chen}}, \bibinfo {author}
  {\bibfnamefont {L.}~\bibnamefont {Yu}}, \bibinfo {author} {\bibfnamefont
  {G.}~\bibnamefont {Liu}}, \bibinfo {author} {\bibfnamefont {X.}~\bibnamefont
  {Dong}}, \bibinfo {author} {\bibfnamefont {J.}~\bibnamefont {Zhang}},
  \bibinfo {author} {\bibfnamefont {C.}~\bibnamefont {Chen}}, \bibinfo {author}
  {\bibfnamefont {Z.}~\bibnamefont {Xu}}, \bibinfo {author} {\bibfnamefont
  {X.}~\bibnamefont {Chen}}, \bibinfo {author} {\bibfnamefont {X.}~\bibnamefont
  {Ma}}, \bibinfo {author} {\bibfnamefont {Q.}~\bibnamefont {Xue}},\ and\
  \bibinfo {author} {\bibfnamefont {X.~J.}\ \bibnamefont {Zhou}},\ }\bibfield
  {title} {\bibinfo {title} {Phase diagram and electronic indication of
  high-temperature superconductivity at 65{\hspace{0.167em}}k in single-layer
  {FeSe} films},\ }\href {https://doi.org/10.1038/nmat3648} {\bibfield
  {journal} {\bibinfo  {journal} {Nature Materials}\ }\textbf {\bibinfo
  {volume} {12}},\ \bibinfo {pages} {605} (\bibinfo {year} {2013})}\BibitemShut
  {NoStop}%
\bibitem [{\citenamefont {Bozovic}\ and\ \citenamefont
  {Ahn}(2014)}]{Bozovic2014}%
  \BibitemOpen
  \bibfield  {author} {\bibinfo {author} {\bibfnamefont {I.}~\bibnamefont
  {Bozovic}}\ and\ \bibinfo {author} {\bibfnamefont {C.}~\bibnamefont {Ahn}},\
  }\bibfield  {title} {\bibinfo {title} {A new frontier for
  superconductivity},\ }\href {https://doi.org/10.1038/nphys3177} {\bibfield
  {journal} {\bibinfo  {journal} {Nature Physics}\ }\textbf {\bibinfo {volume}
  {10}},\ \bibinfo {pages} {892} (\bibinfo {year} {2014})}\BibitemShut
  {NoStop}%
\bibitem [{\citenamefont {Ge}\ \emph {et~al.}(2014)\citenamefont {Ge},
  \citenamefont {Liu}, \citenamefont {Liu}, \citenamefont {Gao}, \citenamefont
  {Qian}, \citenamefont {Xue}, \citenamefont {Liu},\ and\ \citenamefont
  {Jia}}]{Ge2014}%
  \BibitemOpen
  \bibfield  {author} {\bibinfo {author} {\bibfnamefont {J.-F.}\ \bibnamefont
  {Ge}}, \bibinfo {author} {\bibfnamefont {Z.-L.}\ \bibnamefont {Liu}},
  \bibinfo {author} {\bibfnamefont {C.}~\bibnamefont {Liu}}, \bibinfo {author}
  {\bibfnamefont {C.-L.}\ \bibnamefont {Gao}}, \bibinfo {author} {\bibfnamefont
  {D.}~\bibnamefont {Qian}}, \bibinfo {author} {\bibfnamefont {Q.-K.}\
  \bibnamefont {Xue}}, \bibinfo {author} {\bibfnamefont {Y.}~\bibnamefont
  {Liu}},\ and\ \bibinfo {author} {\bibfnamefont {J.-F.}\ \bibnamefont {Jia}},\
  }\bibfield  {title} {\bibinfo {title} {Superconductivity above 100 {K} in
  single-layer {FeSe} films on doped {SrTiO$_3$}},\ }\href
  {https://doi.org/10.1038/nmat4153} {\bibfield  {journal} {\bibinfo  {journal}
  {Nature Materials}\ }\textbf {\bibinfo {volume} {14}},\ \bibinfo {pages}
  {285} (\bibinfo {year} {2014})}\BibitemShut {NoStop}%
\bibitem [{\citenamefont {Lee}\ \emph {et~al.}(2014)\citenamefont {Lee},
  \citenamefont {Schmitt}, \citenamefont {Moore}, \citenamefont {Johnston},
  \citenamefont {Cui}, \citenamefont {Li}, \citenamefont {Yi}, \citenamefont
  {Liu}, \citenamefont {Hashimoto}, \citenamefont {Zhang}, \citenamefont {Lu},
  \citenamefont {Devereaux}, \citenamefont {Lee},\ and\ \citenamefont
  {Shen}}]{Lee2014}%
  \BibitemOpen
  \bibfield  {author} {\bibinfo {author} {\bibfnamefont {J.~J.}\ \bibnamefont
  {Lee}}, \bibinfo {author} {\bibfnamefont {F.~T.}\ \bibnamefont {Schmitt}},
  \bibinfo {author} {\bibfnamefont {R.~G.}\ \bibnamefont {Moore}}, \bibinfo
  {author} {\bibfnamefont {S.}~\bibnamefont {Johnston}}, \bibinfo {author}
  {\bibfnamefont {Y.-T.}\ \bibnamefont {Cui}}, \bibinfo {author} {\bibfnamefont
  {W.}~\bibnamefont {Li}}, \bibinfo {author} {\bibfnamefont {M.}~\bibnamefont
  {Yi}}, \bibinfo {author} {\bibfnamefont {Z.~K.}\ \bibnamefont {Liu}},
  \bibinfo {author} {\bibfnamefont {M.}~\bibnamefont {Hashimoto}}, \bibinfo
  {author} {\bibfnamefont {Y.}~\bibnamefont {Zhang}}, \bibinfo {author}
  {\bibfnamefont {D.~H.}\ \bibnamefont {Lu}}, \bibinfo {author} {\bibfnamefont
  {T.~P.}\ \bibnamefont {Devereaux}}, \bibinfo {author} {\bibfnamefont {D.-H.}\
  \bibnamefont {Lee}},\ and\ \bibinfo {author} {\bibfnamefont {Z.-X.}\
  \bibnamefont {Shen}},\ }\bibfield  {title} {\bibinfo {title} {Interfacial
  mode coupling as the origin of the enhancement of {T$_c$} in {FeSe} films on
  {SrTiO$_3$}},\ }\href {http://dx.doi.org/10.1038/nature13894} {\bibfield
  {journal} {\bibinfo  {journal} {Nature}\ }\textbf {\bibinfo {volume} {515}},\
  \bibinfo {pages} {245} (\bibinfo {year} {2014})}\BibitemShut {NoStop}%
\bibitem [{\citenamefont {Peng}\ \emph {et~al.}(2014)\citenamefont {Peng},
  \citenamefont {Xu}, \citenamefont {Tan}, \citenamefont {Cao}, \citenamefont
  {Xia}, \citenamefont {Shen}, \citenamefont {Huang}, \citenamefont {Wen},
  \citenamefont {Song}, \citenamefont {Zhang}, \citenamefont {Xie},
  \citenamefont {Gong},\ and\ \citenamefont {Feng}}]{Peng2014}%
  \BibitemOpen
  \bibfield  {author} {\bibinfo {author} {\bibfnamefont {R.}~\bibnamefont
  {Peng}}, \bibinfo {author} {\bibfnamefont {H.~C.}\ \bibnamefont {Xu}},
  \bibinfo {author} {\bibfnamefont {S.~Y.}\ \bibnamefont {Tan}}, \bibinfo
  {author} {\bibfnamefont {H.~Y.}\ \bibnamefont {Cao}}, \bibinfo {author}
  {\bibfnamefont {M.}~\bibnamefont {Xia}}, \bibinfo {author} {\bibfnamefont
  {X.~P.}\ \bibnamefont {Shen}}, \bibinfo {author} {\bibfnamefont {Z.~C.}\
  \bibnamefont {Huang}}, \bibinfo {author} {\bibfnamefont {C.}~\bibnamefont
  {Wen}}, \bibinfo {author} {\bibfnamefont {Q.}~\bibnamefont {Song}}, \bibinfo
  {author} {\bibfnamefont {T.}~\bibnamefont {Zhang}}, \bibinfo {author}
  {\bibfnamefont {B.~P.}\ \bibnamefont {Xie}}, \bibinfo {author} {\bibfnamefont
  {X.~G.}\ \bibnamefont {Gong}},\ and\ \bibinfo {author} {\bibfnamefont
  {D.~L.}\ \bibnamefont {Feng}},\ }\bibfield  {title} {\bibinfo {title} {Tuning
  the band structure and superconductivity in single-layer {FeSe} by interface
  engineering},\ }\bibfield  {journal} {\bibinfo  {journal} {Nature
  Communications}\ }\textbf {\bibinfo {volume} {5}},\ \href
  {https://doi.org/10.1038/ncomms6044} {10.1038/ncomms6044} (\bibinfo {year}
  {2014})\BibitemShut {NoStop}%
\bibitem [{\citenamefont {Zhang}\ \emph
  {et~al.}(2017{\natexlab{a}})\citenamefont {Zhang}, \citenamefont {Liu},
  \citenamefont {Chen}, \citenamefont {Xie}, \citenamefont {He}, \citenamefont
  {Tang}, \citenamefont {He}, \citenamefont {Li}, \citenamefont {Jia},
  \citenamefont {Rebec}, \citenamefont {Ma}, \citenamefont {Yan}, \citenamefont
  {Hashimoto}, \citenamefont {Lu}, \citenamefont {Mo}, \citenamefont {Hikita},
  \citenamefont {Moore}, \citenamefont {Hwang}, \citenamefont {Lee},\ and\
  \citenamefont {Shen}}]{Zhang2017}%
  \BibitemOpen
  \bibfield  {author} {\bibinfo {author} {\bibfnamefont {C.}~\bibnamefont
  {Zhang}}, \bibinfo {author} {\bibfnamefont {Z.}~\bibnamefont {Liu}}, \bibinfo
  {author} {\bibfnamefont {Z.}~\bibnamefont {Chen}}, \bibinfo {author}
  {\bibfnamefont {Y.}~\bibnamefont {Xie}}, \bibinfo {author} {\bibfnamefont
  {R.}~\bibnamefont {He}}, \bibinfo {author} {\bibfnamefont {S.}~\bibnamefont
  {Tang}}, \bibinfo {author} {\bibfnamefont {J.}~\bibnamefont {He}}, \bibinfo
  {author} {\bibfnamefont {W.}~\bibnamefont {Li}}, \bibinfo {author}
  {\bibfnamefont {T.}~\bibnamefont {Jia}}, \bibinfo {author} {\bibfnamefont
  {S.~N.}\ \bibnamefont {Rebec}}, \bibinfo {author} {\bibfnamefont {E.~Y.}\
  \bibnamefont {Ma}}, \bibinfo {author} {\bibfnamefont {H.}~\bibnamefont
  {Yan}}, \bibinfo {author} {\bibfnamefont {M.}~\bibnamefont {Hashimoto}},
  \bibinfo {author} {\bibfnamefont {D.}~\bibnamefont {Lu}}, \bibinfo {author}
  {\bibfnamefont {S.-K.}\ \bibnamefont {Mo}}, \bibinfo {author} {\bibfnamefont
  {Y.}~\bibnamefont {Hikita}}, \bibinfo {author} {\bibfnamefont {R.~G.}\
  \bibnamefont {Moore}}, \bibinfo {author} {\bibfnamefont {H.~Y.}\ \bibnamefont
  {Hwang}}, \bibinfo {author} {\bibfnamefont {D.}~\bibnamefont {Lee}},\ and\
  \bibinfo {author} {\bibfnamefont {Z.}~\bibnamefont {Shen}},\ }\bibfield
  {title} {\bibinfo {title} {Ubiquitous strong electron{\textendash}phonon
  coupling at the interface of {FeSe}/{SrTiO$_3$}},\ }\bibfield  {journal}
  {\bibinfo  {journal} {Nature Communications}\ }\textbf {\bibinfo {volume}
  {8}},\ \href {https://doi.org/10.1038/ncomms14468} {10.1038/ncomms14468}
  (\bibinfo {year} {2017}{\natexlab{a}})\BibitemShut {NoStop}%
\bibitem [{\citenamefont {Zhao}\ \emph {et~al.}(2018)\citenamefont {Zhao},
  \citenamefont {Li}, \citenamefont {Chang}, \citenamefont {Jiang},
  \citenamefont {Wu}, \citenamefont {Liu}, \citenamefont {Moodera},
  \citenamefont {Zhu},\ and\ \citenamefont {Chan}}]{Zhao2018}%
  \BibitemOpen
  \bibfield  {author} {\bibinfo {author} {\bibfnamefont {W.}~\bibnamefont
  {Zhao}}, \bibinfo {author} {\bibfnamefont {M.}~\bibnamefont {Li}}, \bibinfo
  {author} {\bibfnamefont {C.-Z.}\ \bibnamefont {Chang}}, \bibinfo {author}
  {\bibfnamefont {J.}~\bibnamefont {Jiang}}, \bibinfo {author} {\bibfnamefont
  {L.}~\bibnamefont {Wu}}, \bibinfo {author} {\bibfnamefont {C.}~\bibnamefont
  {Liu}}, \bibinfo {author} {\bibfnamefont {J.~S.}\ \bibnamefont {Moodera}},
  \bibinfo {author} {\bibfnamefont {Y.}~\bibnamefont {Zhu}},\ and\ \bibinfo
  {author} {\bibfnamefont {M.~H.~W.}\ \bibnamefont {Chan}},\ }\bibfield
  {title} {\bibinfo {title} {Direct imaging of electron transfer and its
  influence on superconducting pairing at {FeSe}/{SrTiO$_3$} interface},\
  }\bibfield  {journal} {\bibinfo  {journal} {Science Advances}\ }\textbf
  {\bibinfo {volume} {4}},\ \href {https://doi.org/10.1126/sciadv.aao2682}
  {10.1126/sciadv.aao2682} (\bibinfo {year} {2018})\BibitemShut {NoStop}%
\bibitem [{\citenamefont {Zhou}\ \emph {et~al.}(2018)\citenamefont {Zhou},
  \citenamefont {Zhang}, \citenamefont {Zheng}, \citenamefont {Zhang},
  \citenamefont {Liu}, \citenamefont {Wang}, \citenamefont {Song},
  \citenamefont {He}, \citenamefont {Ma}, \citenamefont {Gu}, \citenamefont
  {Zhang}, \citenamefont {Wang},\ and\ \citenamefont {Xue}}]{Zhou2018}%
  \BibitemOpen
  \bibfield  {author} {\bibinfo {author} {\bibfnamefont {G.}~\bibnamefont
  {Zhou}}, \bibinfo {author} {\bibfnamefont {Q.}~\bibnamefont {Zhang}},
  \bibinfo {author} {\bibfnamefont {F.}~\bibnamefont {Zheng}}, \bibinfo
  {author} {\bibfnamefont {D.}~\bibnamefont {Zhang}}, \bibinfo {author}
  {\bibfnamefont {C.}~\bibnamefont {Liu}}, \bibinfo {author} {\bibfnamefont
  {X.}~\bibnamefont {Wang}}, \bibinfo {author} {\bibfnamefont {C.-L.}\
  \bibnamefont {Song}}, \bibinfo {author} {\bibfnamefont {K.}~\bibnamefont
  {He}}, \bibinfo {author} {\bibfnamefont {X.-C.}\ \bibnamefont {Ma}}, \bibinfo
  {author} {\bibfnamefont {L.}~\bibnamefont {Gu}}, \bibinfo {author}
  {\bibfnamefont {P.}~\bibnamefont {Zhang}}, \bibinfo {author} {\bibfnamefont
  {L.}~\bibnamefont {Wang}},\ and\ \bibinfo {author} {\bibfnamefont {Q.-K.}\
  \bibnamefont {Xue}},\ }\bibfield  {title} {\bibinfo {title} {Interface
  enhanced superconductivity in monolayer {FeSe} films on {MgO}(001): charge
  transfer with atomic substitution},\ }\href
  {https://doi.org/10.1016/j.scib.2018.05.016} {\bibfield  {journal} {\bibinfo
  {journal} {Science Bulletin}\ }\textbf {\bibinfo {volume} {63}},\ \bibinfo
  {pages} {747} (\bibinfo {year} {2018})}\BibitemShut {NoStop}%
\bibitem [{\citenamefont {Song}\ \emph {et~al.}(2019)\citenamefont {Song},
  \citenamefont {Yu}, \citenamefont {Lou}, \citenamefont {Xie}, \citenamefont
  {Xu}, \citenamefont {Wen}, \citenamefont {Yao}, \citenamefont {Zhang},
  \citenamefont {Zhu}, \citenamefont {Guo}, \citenamefont {Peng},\ and\
  \citenamefont {Feng}}]{Song2019}%
  \BibitemOpen
  \bibfield  {author} {\bibinfo {author} {\bibfnamefont {Q.}~\bibnamefont
  {Song}}, \bibinfo {author} {\bibfnamefont {T.~L.}\ \bibnamefont {Yu}},
  \bibinfo {author} {\bibfnamefont {X.}~\bibnamefont {Lou}}, \bibinfo {author}
  {\bibfnamefont {B.~P.}\ \bibnamefont {Xie}}, \bibinfo {author} {\bibfnamefont
  {H.~C.}\ \bibnamefont {Xu}}, \bibinfo {author} {\bibfnamefont {C.~H.~P.}\
  \bibnamefont {Wen}}, \bibinfo {author} {\bibfnamefont {Q.}~\bibnamefont
  {Yao}}, \bibinfo {author} {\bibfnamefont {S.~Y.}\ \bibnamefont {Zhang}},
  \bibinfo {author} {\bibfnamefont {X.~T.}\ \bibnamefont {Zhu}}, \bibinfo
  {author} {\bibfnamefont {J.~D.}\ \bibnamefont {Guo}}, \bibinfo {author}
  {\bibfnamefont {R.}~\bibnamefont {Peng}},\ and\ \bibinfo {author}
  {\bibfnamefont {D.~L.}\ \bibnamefont {Feng}},\ }\bibfield  {title} {\bibinfo
  {title} {Evidence of cooperative effect on the enhanced superconducting
  transition temperature at the {FeSe}/{SrTiO$_3$} interface},\ }\bibfield
  {journal} {\bibinfo  {journal} {Nature Communications}\ }\textbf {\bibinfo
  {volume} {10}},\ \href {https://doi.org/10.1038/s41467-019-08560-z}
  {10.1038/s41467-019-08560-z} (\bibinfo {year} {2019})\BibitemShut {NoStop}%
\bibitem [{\citenamefont {Xu}\ \emph {et~al.}(2020)\citenamefont {Xu},
  \citenamefont {Zhang}, \citenamefont {Zhu},\ and\ \citenamefont
  {Guo}}]{Xu2020}%
  \BibitemOpen
  \bibfield  {author} {\bibinfo {author} {\bibfnamefont {X.}~\bibnamefont
  {Xu}}, \bibinfo {author} {\bibfnamefont {S.}~\bibnamefont {Zhang}}, \bibinfo
  {author} {\bibfnamefont {X.}~\bibnamefont {Zhu}},\ and\ \bibinfo {author}
  {\bibfnamefont {J.}~\bibnamefont {Guo}},\ }\bibfield  {title} {\bibinfo
  {title} {Superconductivity enhancement in {FeSe}/{SrTiO$_3$}: a review from
  the perspective of electron--phonon coupling},\ }\href
  {https://doi.org/10.1088/1361-648x/ab85f0} {\bibfield  {journal} {\bibinfo
  {journal} {Journal of Physics: Condensed Matter}\ }\textbf {\bibinfo {volume}
  {32}},\ \bibinfo {pages} {343003} (\bibinfo {year} {2020})}\BibitemShut
  {NoStop}%
\bibitem [{\citenamefont {Seo}\ \emph {et~al.}(2016)\citenamefont {Seo},
  \citenamefont {Kim}, \citenamefont {Kim}, \citenamefont {Jeong},
  \citenamefont {Ok}, \citenamefont {Kim}, \citenamefont {Denlinger},
  \citenamefont {Mo}, \citenamefont {Kim},\ and\ \citenamefont
  {Kim}}]{Seo2016}%
  \BibitemOpen
  \bibfield  {author} {\bibinfo {author} {\bibfnamefont {J.~J.}\ \bibnamefont
  {Seo}}, \bibinfo {author} {\bibfnamefont {B.~Y.}\ \bibnamefont {Kim}},
  \bibinfo {author} {\bibfnamefont {B.~S.}\ \bibnamefont {Kim}}, \bibinfo
  {author} {\bibfnamefont {J.~K.}\ \bibnamefont {Jeong}}, \bibinfo {author}
  {\bibfnamefont {J.~M.}\ \bibnamefont {Ok}}, \bibinfo {author} {\bibfnamefont
  {J.~S.}\ \bibnamefont {Kim}}, \bibinfo {author} {\bibfnamefont {J.~D.}\
  \bibnamefont {Denlinger}}, \bibinfo {author} {\bibfnamefont {S.~K.}\
  \bibnamefont {Mo}}, \bibinfo {author} {\bibfnamefont {C.}~\bibnamefont
  {Kim}},\ and\ \bibinfo {author} {\bibfnamefont {Y.~K.}\ \bibnamefont {Kim}},\
  }\bibfield  {title} {\bibinfo {title} {Superconductivity below
  20{\hspace{0.167em}}k in heavily electron-doped surface layer of {FeSe} bulk
  crystal},\ }\bibfield  {journal} {\bibinfo  {journal} {Nature
  Communications}\ }\textbf {\bibinfo {volume} {7}},\ \href
  {https://doi.org/10.1038/ncomms11116} {10.1038/ncomms11116} (\bibinfo {year}
  {2016})\BibitemShut {NoStop}%
\bibitem [{\citenamefont {Zhang}\ \emph
  {et~al.}(2017{\natexlab{b}})\citenamefont {Zhang}, \citenamefont {Zhang},
  \citenamefont {Lu}, \citenamefont {Liu}, \citenamefont {Zhou}, \citenamefont
  {Ma}, \citenamefont {Wang}, \citenamefont {Jiang}, \citenamefont {Xue},\ and\
  \citenamefont {Bao}}]{Zhang2017a}%
  \BibitemOpen
  \bibfield  {author} {\bibinfo {author} {\bibfnamefont {H.}~\bibnamefont
  {Zhang}}, \bibinfo {author} {\bibfnamefont {D.}~\bibnamefont {Zhang}},
  \bibinfo {author} {\bibfnamefont {X.}~\bibnamefont {Lu}}, \bibinfo {author}
  {\bibfnamefont {C.}~\bibnamefont {Liu}}, \bibinfo {author} {\bibfnamefont
  {G.}~\bibnamefont {Zhou}}, \bibinfo {author} {\bibfnamefont {X.}~\bibnamefont
  {Ma}}, \bibinfo {author} {\bibfnamefont {L.}~\bibnamefont {Wang}}, \bibinfo
  {author} {\bibfnamefont {P.}~\bibnamefont {Jiang}}, \bibinfo {author}
  {\bibfnamefont {Q.-K.}\ \bibnamefont {Xue}},\ and\ \bibinfo {author}
  {\bibfnamefont {X.}~\bibnamefont {Bao}},\ }\bibfield  {title} {\bibinfo
  {title} {Origin of charge transfer and enhanced electron{\textendash}phonon
  coupling in single unit-cell {FeSe} films on {SrTiO$_3$}},\ }\bibfield
  {journal} {\bibinfo  {journal} {Nature Communications}\ }\textbf {\bibinfo
  {volume} {8}},\ \href {https://doi.org/10.1038/s41467-017-00281-5}
  {10.1038/s41467-017-00281-5} (\bibinfo {year}
  {2017}{\natexlab{b}})\BibitemShut {NoStop}%
\bibitem [{\citenamefont {Shi}\ \emph {et~al.}(2017)\citenamefont {Shi},
  \citenamefont {Han}, \citenamefont {Peng}, \citenamefont {Richard},
  \citenamefont {Qian}, \citenamefont {Wu}, \citenamefont {Qiu}, \citenamefont
  {Wang}, \citenamefont {Hu}, \citenamefont {Sun},\ and\ \citenamefont
  {Ding}}]{Shi2017}%
  \BibitemOpen
  \bibfield  {author} {\bibinfo {author} {\bibfnamefont {X.}~\bibnamefont
  {Shi}}, \bibinfo {author} {\bibfnamefont {Z.-Q.}\ \bibnamefont {Han}},
  \bibinfo {author} {\bibfnamefont {X.-L.}\ \bibnamefont {Peng}}, \bibinfo
  {author} {\bibfnamefont {P.}~\bibnamefont {Richard}}, \bibinfo {author}
  {\bibfnamefont {T.}~\bibnamefont {Qian}}, \bibinfo {author} {\bibfnamefont
  {X.-X.}\ \bibnamefont {Wu}}, \bibinfo {author} {\bibfnamefont {M.-W.}\
  \bibnamefont {Qiu}}, \bibinfo {author} {\bibfnamefont {S.~C.}\ \bibnamefont
  {Wang}}, \bibinfo {author} {\bibfnamefont {J.~P.}\ \bibnamefont {Hu}},
  \bibinfo {author} {\bibfnamefont {Y.-J.}\ \bibnamefont {Sun}},\ and\ \bibinfo
  {author} {\bibfnamefont {H.}~\bibnamefont {Ding}},\ }\bibfield  {title}
  {\bibinfo {title} {Enhanced superconductivity accompanying a lifshitz
  transition in electron-doped {FeSe} monolayer},\ }\bibfield  {journal}
  {\bibinfo  {journal} {Nature Communications}\ }\textbf {\bibinfo {volume}
  {8}},\ \href {https://doi.org/10.1038/ncomms14988} {10.1038/ncomms14988}
  (\bibinfo {year} {2017})\BibitemShut {NoStop}%
\bibitem [{\citenamefont {Zhou}\ and\ \citenamefont {Millis}(2016)}]{Zhou2016}%
  \BibitemOpen
  \bibfield  {author} {\bibinfo {author} {\bibfnamefont {Y.}~\bibnamefont
  {Zhou}}\ and\ \bibinfo {author} {\bibfnamefont {A.~J.}\ \bibnamefont
  {Millis}},\ }\bibfield  {title} {\bibinfo {title} {Charge transfer and
  electron-phonon coupling in monolayer {FeSe} on nb-doped {SrTiO$_3$}},\
  }\href {https://doi.org/10.1103/physrevb.93.224506} {\bibfield  {journal}
  {\bibinfo  {journal} {Physical Review B}\ }\textbf {\bibinfo {volume} {93}},\
  \bibinfo {pages} {224506} (\bibinfo {year} {2016})}\BibitemShut {NoStop}%
\bibitem [{\citenamefont {Ibach}\ and\ \citenamefont
  {Mills}(1982)}]{Ibach1982}%
  \BibitemOpen
  \bibfield  {author} {\bibinfo {author} {\bibfnamefont {H.}~\bibnamefont
  {Ibach}}\ and\ \bibinfo {author} {\bibfnamefont {D.}~\bibnamefont {Mills}},\
  }\href@noop {} {\emph {\bibinfo {title} {Electron Energy Loss Spectroscopy
  and Surface Vibrations}}}\ (\bibinfo  {publisher} {Academic},\ \bibinfo
  {address} {New York},\ \bibinfo {year} {1982})\ pp.\ \bibinfo {pages}
  {105--120}\BibitemShut {NoStop}%
\bibitem [{\citenamefont {Ritz}\ and\ \citenamefont {Lüth}(1984)}]{Ritz1984}%
  \BibitemOpen
  \bibfield  {author} {\bibinfo {author} {\bibfnamefont {A.}~\bibnamefont
  {Ritz}}\ and\ \bibinfo {author} {\bibfnamefont {H.}~\bibnamefont {Lüth}},\
  }\bibfield  {title} {\bibinfo {title} {Experimental evidence for surface
  quenching of the surface plasmon on {InSb}(110)},\ }\href
  {https://doi.org/10.1103/physrevlett.52.1242} {\bibfield  {journal} {\bibinfo
   {journal} {Physical Review Letters}\ }\textbf {\bibinfo {volume} {52}},\
  \bibinfo {pages} {1242} (\bibinfo {year} {1984})}\BibitemShut {NoStop}%
\bibitem [{\citenamefont {Schaich}(1984)}]{Schaich1984}%
  \BibitemOpen
  \bibfield  {author} {\bibinfo {author} {\bibfnamefont {W.~L.}\ \bibnamefont
  {Schaich}},\ }\bibfield  {title} {\bibinfo {title} {Surface quenching or
  surface depletion?},\ }\href {https://doi.org/10.1103/physrevlett.53.2059}
  {\bibfield  {journal} {\bibinfo  {journal} {Physical Review Letters}\
  }\textbf {\bibinfo {volume} {53}},\ \bibinfo {pages} {2059} (\bibinfo {year}
  {1984})}\BibitemShut {NoStop}%
\bibitem [{\citenamefont {Lüth}(1988)}]{Lueth1988}%
  \BibitemOpen
  \bibfield  {author} {\bibinfo {author} {\bibfnamefont {H.}~\bibnamefont
  {Lüth}},\ }\bibfield  {title} {\bibinfo {title} {Electron energy loss
  spectroscopy applied to semiconductor space charge layers},\ }\href
  {https://doi.org/10.1016/0042-207x(88)90049-8} {\bibfield  {journal}
  {\bibinfo  {journal} {Vacuum}\ }\textbf {\bibinfo {volume} {38}},\ \bibinfo
  {pages} {223} (\bibinfo {year} {1988})}\BibitemShut {NoStop}%
\bibitem [{\citenamefont {Vig}\ \emph {et~al.}(2017)\citenamefont {Vig},
  \citenamefont {Kogar}, \citenamefont {Mitrano}, \citenamefont {Husain},
  \citenamefont {Venema}, \citenamefont {Rak}, \citenamefont {Mishra},
  \citenamefont {Johnson}, \citenamefont {Gu}, \citenamefont {Fradkin},
  \citenamefont {Norman},\ and\ \citenamefont {Abbamonte}}]{Vig2017}%
  \BibitemOpen
  \bibfield  {author} {\bibinfo {author} {\bibfnamefont {S.}~\bibnamefont
  {Vig}}, \bibinfo {author} {\bibfnamefont {A.}~\bibnamefont {Kogar}}, \bibinfo
  {author} {\bibfnamefont {M.}~\bibnamefont {Mitrano}}, \bibinfo {author}
  {\bibfnamefont {A.}~\bibnamefont {Husain}}, \bibinfo {author} {\bibfnamefont
  {L.}~\bibnamefont {Venema}}, \bibinfo {author} {\bibfnamefont
  {M.}~\bibnamefont {Rak}}, \bibinfo {author} {\bibfnamefont {V.}~\bibnamefont
  {Mishra}}, \bibinfo {author} {\bibfnamefont {P.}~\bibnamefont {Johnson}},
  \bibinfo {author} {\bibfnamefont {G.}~\bibnamefont {Gu}}, \bibinfo {author}
  {\bibfnamefont {E.}~\bibnamefont {Fradkin}}, \bibinfo {author} {\bibfnamefont
  {M.}~\bibnamefont {Norman}},\ and\ \bibinfo {author} {\bibfnamefont
  {P.}~\bibnamefont {Abbamonte}},\ }\bibfield  {title} {\bibinfo {title}
  {Measurement of the dynamic charge response of materials using low-energy,
  momentum-resolved electron energy-loss spectroscopy (m-{EELS})},\ }\bibfield
  {journal} {\bibinfo  {journal} {{SciPost} Physics}\ }\textbf {\bibinfo
  {volume} {3}},\ \href {https://doi.org/10.21468/scipostphys.3.4.026}
  {10.21468/scipostphys.3.4.026} (\bibinfo {year} {2017})\BibitemShut {NoStop}%
\bibitem [{\citenamefont {Husain}\ \emph {et~al.}(2019)\citenamefont {Husain},
  \citenamefont {Mitrano}, \citenamefont {Rak}, \citenamefont {Rubeck},
  \citenamefont {Uchoa}, \citenamefont {March}, \citenamefont {Dwyer},
  \citenamefont {Schneeloch}, \citenamefont {Zhong}, \citenamefont {Gu},\ and\
  \citenamefont {Abbamonte}}]{Husain2019}%
  \BibitemOpen
  \bibfield  {author} {\bibinfo {author} {\bibfnamefont {A.~A.}\ \bibnamefont
  {Husain}}, \bibinfo {author} {\bibfnamefont {M.}~\bibnamefont {Mitrano}},
  \bibinfo {author} {\bibfnamefont {M.~S.}\ \bibnamefont {Rak}}, \bibinfo
  {author} {\bibfnamefont {S.}~\bibnamefont {Rubeck}}, \bibinfo {author}
  {\bibfnamefont {B.}~\bibnamefont {Uchoa}}, \bibinfo {author} {\bibfnamefont
  {K.}~\bibnamefont {March}}, \bibinfo {author} {\bibfnamefont
  {C.}~\bibnamefont {Dwyer}}, \bibinfo {author} {\bibfnamefont
  {J.}~\bibnamefont {Schneeloch}}, \bibinfo {author} {\bibfnamefont
  {R.}~\bibnamefont {Zhong}}, \bibinfo {author} {\bibfnamefont
  {G.}~\bibnamefont {Gu}},\ and\ \bibinfo {author} {\bibfnamefont
  {P.}~\bibnamefont {Abbamonte}},\ }\bibfield  {title} {\bibinfo {title}
  {Crossover of charge fluctuations across the strange metal phase diagram},\
  }\href {https://doi.org/10.1103/physrevx.9.041062} {\bibfield  {journal}
  {\bibinfo  {journal} {Physical Review X}\ }\textbf {\bibinfo {volume} {9}},\
  \bibinfo {pages} {041062} (\bibinfo {year} {2019})}\BibitemShut {NoStop}%
\bibitem [{\citenamefont {Gnezdilov}\ \emph {et~al.}(2013)\citenamefont
  {Gnezdilov}, \citenamefont {Pashkevich}, \citenamefont {Lemmens},
  \citenamefont {Wulferding}, \citenamefont {Shevtsova}, \citenamefont {Gusev},
  \citenamefont {Chareev},\ and\ \citenamefont {Vasiliev}}]{Gnezdilov2013}%
  \BibitemOpen
  \bibfield  {author} {\bibinfo {author} {\bibfnamefont {V.}~\bibnamefont
  {Gnezdilov}}, \bibinfo {author} {\bibfnamefont {Y.~G.}\ \bibnamefont
  {Pashkevich}}, \bibinfo {author} {\bibfnamefont {P.}~\bibnamefont {Lemmens}},
  \bibinfo {author} {\bibfnamefont {D.}~\bibnamefont {Wulferding}}, \bibinfo
  {author} {\bibfnamefont {T.}~\bibnamefont {Shevtsova}}, \bibinfo {author}
  {\bibfnamefont {A.}~\bibnamefont {Gusev}}, \bibinfo {author} {\bibfnamefont
  {D.}~\bibnamefont {Chareev}},\ and\ \bibinfo {author} {\bibfnamefont
  {A.}~\bibnamefont {Vasiliev}},\ }\bibfield  {title} {\bibinfo {title}
  {Interplay between lattice and spin states degree of freedom in the fese
  superconductor: {D}ynamic spin state instabilities},\ }\href
  {https://doi.org/10.1103/PhysRevB.87.144508} {\bibfield  {journal} {\bibinfo
  {journal} {Phys. Rev. B}\ }\textbf {\bibinfo {volume} {87}},\ \bibinfo
  {pages} {144508} (\bibinfo {year} {2013})}\BibitemShut {NoStop}%
\bibitem [{\citenamefont {Zakeri}\ \emph {et~al.}(2017)\citenamefont {Zakeri},
  \citenamefont {Engelhardt}, \citenamefont {Wolf},\ and\ \citenamefont
  {Tacon}}]{Zakeri2017}%
  \BibitemOpen
  \bibfield  {author} {\bibinfo {author} {\bibfnamefont {K.}~\bibnamefont
  {Zakeri}}, \bibinfo {author} {\bibfnamefont {T.}~\bibnamefont {Engelhardt}},
  \bibinfo {author} {\bibfnamefont {T.}~\bibnamefont {Wolf}},\ and\ \bibinfo
  {author} {\bibfnamefont {M.~L.}\ \bibnamefont {Tacon}},\ }\bibfield  {title}
  {\bibinfo {title} {Phonon dispersion relation of single-crystalline
  $\beta-${FeSe}},\ }\href {https://doi.org/10.1103/physrevb.96.094531}
  {\bibfield  {journal} {\bibinfo  {journal} {Physical Review B}\ }\textbf
  {\bibinfo {volume} {96}},\ \bibinfo {pages} {094531} (\bibinfo {year}
  {2017})}\BibitemShut {NoStop}%
\bibitem [{\citenamefont {Zakeri}\ \emph {et~al.}(2018)\citenamefont {Zakeri},
  \citenamefont {Engelhardt}, \citenamefont {Tacon},\ and\ \citenamefont
  {Wolf}}]{Zakeri2018}%
  \BibitemOpen
  \bibfield  {author} {\bibinfo {author} {\bibfnamefont {K.}~\bibnamefont
  {Zakeri}}, \bibinfo {author} {\bibfnamefont {T.}~\bibnamefont {Engelhardt}},
  \bibinfo {author} {\bibfnamefont {M.~L.}\ \bibnamefont {Tacon}},\ and\
  \bibinfo {author} {\bibfnamefont {T.}~\bibnamefont {Wolf}},\ }\bibfield
  {title} {\bibinfo {title} {Phonon spectrum of single-crystalline {FeSe}
  probed by high-resolution electron energy-loss spectroscopy},\ }\href
  {https://doi.org/10.1016/j.physc.2018.02.042} {\bibfield  {journal} {\bibinfo
   {journal} {Physica C: Superconductivity and its Applications}\ }\textbf
  {\bibinfo {volume} {549}},\ \bibinfo {pages} {18} (\bibinfo {year}
  {2018})}\BibitemShut {NoStop}%
\bibitem [{\citenamefont {Zhang}\ \emph {et~al.}(2018)\citenamefont {Zhang},
  \citenamefont {Guan}, \citenamefont {Wang}, \citenamefont {Berlijn},
  \citenamefont {Johnston}, \citenamefont {Jia}, \citenamefont {Liu},
  \citenamefont {Zhu}, \citenamefont {An}, \citenamefont {Xue}, \citenamefont
  {Cao}, \citenamefont {Yang}, \citenamefont {Wang}, \citenamefont {Zhang},
  \citenamefont {Plummer}, \citenamefont {Zhu},\ and\ \citenamefont
  {Guo}}]{Zhang2018}%
  \BibitemOpen
  \bibfield  {author} {\bibinfo {author} {\bibfnamefont {S.}~\bibnamefont
  {Zhang}}, \bibinfo {author} {\bibfnamefont {J.}~\bibnamefont {Guan}},
  \bibinfo {author} {\bibfnamefont {Y.}~\bibnamefont {Wang}}, \bibinfo {author}
  {\bibfnamefont {T.}~\bibnamefont {Berlijn}}, \bibinfo {author} {\bibfnamefont
  {S.}~\bibnamefont {Johnston}}, \bibinfo {author} {\bibfnamefont
  {X.}~\bibnamefont {Jia}}, \bibinfo {author} {\bibfnamefont {B.}~\bibnamefont
  {Liu}}, \bibinfo {author} {\bibfnamefont {Q.}~\bibnamefont {Zhu}}, \bibinfo
  {author} {\bibfnamefont {Q.}~\bibnamefont {An}}, \bibinfo {author}
  {\bibfnamefont {S.}~\bibnamefont {Xue}}, \bibinfo {author} {\bibfnamefont
  {Y.}~\bibnamefont {Cao}}, \bibinfo {author} {\bibfnamefont {F.}~\bibnamefont
  {Yang}}, \bibinfo {author} {\bibfnamefont {W.}~\bibnamefont {Wang}}, \bibinfo
  {author} {\bibfnamefont {J.}~\bibnamefont {Zhang}}, \bibinfo {author}
  {\bibfnamefont {E.~W.}\ \bibnamefont {Plummer}}, \bibinfo {author}
  {\bibfnamefont {X.}~\bibnamefont {Zhu}},\ and\ \bibinfo {author}
  {\bibfnamefont {J.}~\bibnamefont {Guo}},\ }\bibfield  {title} {\bibinfo
  {title} {Lattice dynamics of ultrathin {FeSe} films on {SrTiO$_3$}},\ }\href
  {https://doi.org/10.1103/physrevb.97.035408} {\bibfield  {journal} {\bibinfo
  {journal} {Physical Review B}\ }\textbf {\bibinfo {volume} {97}},\ \bibinfo
  {pages} {035408} (\bibinfo {year} {2018})}\BibitemShut {NoStop}%
\bibitem [{\citenamefont {Conard}\ \emph {et~al.}(1993)\citenamefont {Conard},
  \citenamefont {Philippe}, \citenamefont {Thiry}, \citenamefont {Lambin},\
  and\ \citenamefont {Caudano}}]{Conard1993}%
  \BibitemOpen
  \bibfield  {author} {\bibinfo {author} {\bibfnamefont {T.}~\bibnamefont
  {Conard}}, \bibinfo {author} {\bibfnamefont {L.}~\bibnamefont {Philippe}},
  \bibinfo {author} {\bibfnamefont {P.}~\bibnamefont {Thiry}}, \bibinfo
  {author} {\bibfnamefont {P.}~\bibnamefont {Lambin}},\ and\ \bibinfo {author}
  {\bibfnamefont {R.}~\bibnamefont {Caudano}},\ }\bibfield  {title} {\bibinfo
  {title} {Electron energy-loss spectroscopy and dynamics of
  {SrTiO$_3$}(100)},\ }\href {https://doi.org/10.1016/0167-2584(93)90456-s}
  {\bibfield  {journal} {\bibinfo  {journal} {Surface Science Letters}\
  }\textbf {\bibinfo {volume} {287-288}},\ \bibinfo {pages} {A390} (\bibinfo
  {year} {1993})}\BibitemShut {NoStop}%
\bibitem [{\citenamefont {Ibach}(1970)}]{Ibach1970}%
  \BibitemOpen
  \bibfield  {author} {\bibinfo {author} {\bibfnamefont {H.}~\bibnamefont
  {Ibach}},\ }\bibfield  {title} {\bibinfo {title} {Optical surface phonons in
  zinc oxide detected by slow-electron spectroscopy},\ }\href
  {https://doi.org/10.1103/physrevlett.24.1416} {\bibfield  {journal} {\bibinfo
   {journal} {Physical Review Letters}\ }\textbf {\bibinfo {volume} {24}},\
  \bibinfo {pages} {1416} (\bibinfo {year} {1970})}\BibitemShut {NoStop}%
\bibitem [{\citenamefont {Peng}\ \emph {et~al.}(2020)\citenamefont {Peng},
  \citenamefont {Zou}, \citenamefont {Han}, \citenamefont {Albright},
  \citenamefont {Hong}, \citenamefont {Lau}, \citenamefont {Xu}, \citenamefont
  {Zhu}, \citenamefont {Walker},\ and\ \citenamefont {Ahn}}]{Peng2020}%
  \BibitemOpen
  \bibfield  {author} {\bibinfo {author} {\bibfnamefont {R.}~\bibnamefont
  {Peng}}, \bibinfo {author} {\bibfnamefont {K.}~\bibnamefont {Zou}}, \bibinfo
  {author} {\bibfnamefont {M.~G.}\ \bibnamefont {Han}}, \bibinfo {author}
  {\bibfnamefont {S.~D.}\ \bibnamefont {Albright}}, \bibinfo {author}
  {\bibfnamefont {H.}~\bibnamefont {Hong}}, \bibinfo {author} {\bibfnamefont
  {C.}~\bibnamefont {Lau}}, \bibinfo {author} {\bibfnamefont {H.~C.}\
  \bibnamefont {Xu}}, \bibinfo {author} {\bibfnamefont {Y.}~\bibnamefont
  {Zhu}}, \bibinfo {author} {\bibfnamefont {F.~J.}\ \bibnamefont {Walker}},\
  and\ \bibinfo {author} {\bibfnamefont {C.~H.}\ \bibnamefont {Ahn}},\
  }\bibfield  {title} {\bibinfo {title} {Picoscale structural insight into
  superconductivity of monolayer {FeSe}/{SrTiO$_3$}},\ }\bibfield  {journal}
  {\bibinfo  {journal} {Science Advances}\ }\textbf {\bibinfo {volume} {6}},\
  \href {https://doi.org/10.1126/sciadv.aay4517} {10.1126/sciadv.aay4517}
  (\bibinfo {year} {2020})\BibitemShut {NoStop}%
\bibitem [{\citenamefont {Galzerani}\ and\ \citenamefont
  {Katiyar}(1982)}]{Galzerani1982}%
  \BibitemOpen
  \bibfield  {author} {\bibinfo {author} {\bibfnamefont {J.}~\bibnamefont
  {Galzerani}}\ and\ \bibinfo {author} {\bibfnamefont {R.}~\bibnamefont
  {Katiyar}},\ }\bibfield  {title} {\bibinfo {title} {The infrared reflectivity
  in {SrTiO$_3$} and the antidistortive transition},\ }\href
  {https://doi.org/10.1016/0038-1098(82)90189-2} {\bibfield  {journal}
  {\bibinfo  {journal} {Solid State Communications}\ }\textbf {\bibinfo
  {volume} {41}},\ \bibinfo {pages} {515} (\bibinfo {year} {1982})}\BibitemShut
  {NoStop}%
\bibitem [{\citenamefont {Gervais}\ \emph {et~al.}(1993)\citenamefont
  {Gervais}, \citenamefont {Servoin}, \citenamefont {Baratoff}, \citenamefont
  {Bednorz},\ and\ \citenamefont {Binnig}}]{Gervais1993}%
  \BibitemOpen
  \bibfield  {author} {\bibinfo {author} {\bibfnamefont {F.}~\bibnamefont
  {Gervais}}, \bibinfo {author} {\bibfnamefont {J.-L.}\ \bibnamefont
  {Servoin}}, \bibinfo {author} {\bibfnamefont {A.}~\bibnamefont {Baratoff}},
  \bibinfo {author} {\bibfnamefont {J.~G.}\ \bibnamefont {Bednorz}},\ and\
  \bibinfo {author} {\bibfnamefont {G.}~\bibnamefont {Binnig}},\ }\bibfield
  {title} {\bibinfo {title} {Temperature dependence of plasmons in nb-doped
  {SrTiO$_3$}},\ }\href {https://doi.org/10.1103/physrevb.47.8187} {\bibfield
  {journal} {\bibinfo  {journal} {Physical Review B}\ }\textbf {\bibinfo
  {volume} {47}},\ \bibinfo {pages} {8187} (\bibinfo {year}
  {1993})}\BibitemShut {NoStop}%
\bibitem [{\citenamefont {Collignon}\ \emph {et~al.}(2020)\citenamefont
  {Collignon}, \citenamefont {Bourges}, \citenamefont {Fauqu{\'{e}}},\ and\
  \citenamefont {Behnia}}]{Collignon2020}%
  \BibitemOpen
  \bibfield  {author} {\bibinfo {author} {\bibfnamefont {C.}~\bibnamefont
  {Collignon}}, \bibinfo {author} {\bibfnamefont {P.}~\bibnamefont {Bourges}},
  \bibinfo {author} {\bibfnamefont {B.}~\bibnamefont {Fauqu{\'{e}}}},\ and\
  \bibinfo {author} {\bibfnamefont {K.}~\bibnamefont {Behnia}},\ }\bibfield
  {title} {\bibinfo {title} {Heavy nondegenerate electrons in doped strontium
  titanate},\ }\href {https://doi.org/10.1103/physrevx.10.031025} {\bibfield
  {journal} {\bibinfo  {journal} {Physical Review X}\ }\textbf {\bibinfo
  {volume} {10}},\ \bibinfo {pages} {031025} (\bibinfo {year}
  {2020})}\BibitemShut {NoStop}%
\bibitem [{\citenamefont {Zhang}\ \emph {et~al.}(2016)\citenamefont {Zhang},
  \citenamefont {Guan}, \citenamefont {Jia}, \citenamefont {Liu}, \citenamefont
  {Wang}, \citenamefont {Li}, \citenamefont {Wang}, \citenamefont {Ma},
  \citenamefont {Xue}, \citenamefont {Zhang}, \citenamefont {Plummer},
  \citenamefont {Zhu},\ and\ \citenamefont {Guo}}]{Zhang2016}%
  \BibitemOpen
  \bibfield  {author} {\bibinfo {author} {\bibfnamefont {S.}~\bibnamefont
  {Zhang}}, \bibinfo {author} {\bibfnamefont {J.}~\bibnamefont {Guan}},
  \bibinfo {author} {\bibfnamefont {X.}~\bibnamefont {Jia}}, \bibinfo {author}
  {\bibfnamefont {B.}~\bibnamefont {Liu}}, \bibinfo {author} {\bibfnamefont
  {W.}~\bibnamefont {Wang}}, \bibinfo {author} {\bibfnamefont {F.}~\bibnamefont
  {Li}}, \bibinfo {author} {\bibfnamefont {L.}~\bibnamefont {Wang}}, \bibinfo
  {author} {\bibfnamefont {X.}~\bibnamefont {Ma}}, \bibinfo {author}
  {\bibfnamefont {Q.}~\bibnamefont {Xue}}, \bibinfo {author} {\bibfnamefont
  {J.}~\bibnamefont {Zhang}}, \bibinfo {author} {\bibfnamefont {E.~W.}\
  \bibnamefont {Plummer}}, \bibinfo {author} {\bibfnamefont {X.}~\bibnamefont
  {Zhu}},\ and\ \bibinfo {author} {\bibfnamefont {J.}~\bibnamefont {Guo}},\
  }\bibfield  {title} {\bibinfo {title} {Role of {SrTiO}$_3$ phonon penetrating
  into thin {FeSe} films in the enhancement of superconductivity},\ }\href
  {https://doi.org/10.1103/PhysRevB.94.081116} {\bibfield  {journal} {\bibinfo
  {journal} {Phys. Rev. B}\ }\textbf {\bibinfo {volume} {94}},\ \bibinfo
  {pages} {081116} (\bibinfo {year} {2016})}\BibitemShut {NoStop}%
\bibitem [{\citenamefont {Jandke}\ \emph {et~al.}(2019)\citenamefont {Jandke},
  \citenamefont {Yang}, \citenamefont {Hlobil}, \citenamefont {Engelhardt},
  \citenamefont {Rau}, \citenamefont {Zakeri}, \citenamefont {Gao},
  \citenamefont {Schmalian},\ and\ \citenamefont {Wulfhekel}}]{Jandke2019}%
  \BibitemOpen
  \bibfield  {author} {\bibinfo {author} {\bibfnamefont {J.}~\bibnamefont
  {Jandke}}, \bibinfo {author} {\bibfnamefont {F.}~\bibnamefont {Yang}},
  \bibinfo {author} {\bibfnamefont {P.}~\bibnamefont {Hlobil}}, \bibinfo
  {author} {\bibfnamefont {T.}~\bibnamefont {Engelhardt}}, \bibinfo {author}
  {\bibfnamefont {D.}~\bibnamefont {Rau}}, \bibinfo {author} {\bibfnamefont
  {K.}~\bibnamefont {Zakeri}}, \bibinfo {author} {\bibfnamefont
  {C.}~\bibnamefont {Gao}}, \bibinfo {author} {\bibfnamefont {J.}~\bibnamefont
  {Schmalian}},\ and\ \bibinfo {author} {\bibfnamefont {W.}~\bibnamefont
  {Wulfhekel}},\ }\bibfield  {title} {\bibinfo {title} {Unconventional pairing
  in single {FeSe} layers},\ }\href
  {https://doi.org/10.1103/physrevb.100.020503} {\bibfield  {journal} {\bibinfo
   {journal} {Physical Review B}\ }\textbf {\bibinfo {volume} {100}},\ \bibinfo
  {pages} {020503} (\bibinfo {year} {2019})}\BibitemShut {NoStop}%
\bibitem [{\citenamefont {Wang}\ \emph
  {et~al.}(2016{\natexlab{a}})\citenamefont {Wang}, \citenamefont {Linscheid},
  \citenamefont {Berlijn},\ and\ \citenamefont {Johnston}}]{Wang2016}%
  \BibitemOpen
  \bibfield  {author} {\bibinfo {author} {\bibfnamefont {Y.}~\bibnamefont
  {Wang}}, \bibinfo {author} {\bibfnamefont {A.}~\bibnamefont {Linscheid}},
  \bibinfo {author} {\bibfnamefont {T.}~\bibnamefont {Berlijn}},\ and\ \bibinfo
  {author} {\bibfnamefont {S.}~\bibnamefont {Johnston}},\ }\bibfield  {title}
  {\bibinfo {title} {\textit{Ab initio} study of cross-interface
  electron-phonon couplings in {FeSe} thin films on {SrTiO$_3$}},\ }\href
  {https://doi.org/10.1103/physrevb.93.134513} {\bibfield  {journal} {\bibinfo
  {journal} {Physical Review B}\ }\textbf {\bibinfo {volume} {93}},\ \bibinfo
  {pages} {134513} (\bibinfo {year} {2016}{\natexlab{a}})}\BibitemShut
  {NoStop}%
\bibitem [{\citenamefont {Zhou}\ and\ \citenamefont {Millis}(2017)}]{Zhou2017}%
  \BibitemOpen
  \bibfield  {author} {\bibinfo {author} {\bibfnamefont {Y.}~\bibnamefont
  {Zhou}}\ and\ \bibinfo {author} {\bibfnamefont {A.~J.}\ \bibnamefont
  {Millis}},\ }\bibfield  {title} {\bibinfo {title} {Dipolar phonons and
  electronic screening in monolayer {FeSe} on {SrTiO$_3$}},\ }\href
  {https://doi.org/10.1103/physrevb.96.054516} {\bibfield  {journal} {\bibinfo
  {journal} {Physical Review B}\ }\textbf {\bibinfo {volume} {96}},\ \bibinfo
  {pages} {054516} (\bibinfo {year} {2017})}\BibitemShut {NoStop}%
\bibitem [{\citenamefont {Faeth}\ \emph {et~al.}(2021)\citenamefont {Faeth},
  \citenamefont {Xie}, \citenamefont {Yang}, \citenamefont {Kawasaki},
  \citenamefont {Nelson}, \citenamefont {Zhang}, \citenamefont {Parzyck},
  \citenamefont {Mishra}, \citenamefont {Li}, \citenamefont {Jozwiak},
  \citenamefont {Bostwick}, \citenamefont {Rotenberg}, \citenamefont {Schlom},\
  and\ \citenamefont {Shen}}]{Faeth2021}%
  \BibitemOpen
  \bibfield  {author} {\bibinfo {author} {\bibfnamefont {B.~D.}\ \bibnamefont
  {Faeth}}, \bibinfo {author} {\bibfnamefont {S.}~\bibnamefont {Xie}}, \bibinfo
  {author} {\bibfnamefont {S.}~\bibnamefont {Yang}}, \bibinfo {author}
  {\bibfnamefont {J.~K.}\ \bibnamefont {Kawasaki}}, \bibinfo {author}
  {\bibfnamefont {J.~N.}\ \bibnamefont {Nelson}}, \bibinfo {author}
  {\bibfnamefont {S.}~\bibnamefont {Zhang}}, \bibinfo {author} {\bibfnamefont
  {C.}~\bibnamefont {Parzyck}}, \bibinfo {author} {\bibfnamefont
  {P.}~\bibnamefont {Mishra}}, \bibinfo {author} {\bibfnamefont
  {C.}~\bibnamefont {Li}}, \bibinfo {author} {\bibfnamefont {C.}~\bibnamefont
  {Jozwiak}}, \bibinfo {author} {\bibfnamefont {A.}~\bibnamefont {Bostwick}},
  \bibinfo {author} {\bibfnamefont {E.}~\bibnamefont {Rotenberg}}, \bibinfo
  {author} {\bibfnamefont {D.~G.}\ \bibnamefont {Schlom}},\ and\ \bibinfo
  {author} {\bibfnamefont {K.~M.}\ \bibnamefont {Shen}},\ }\bibfield  {title}
  {\bibinfo {title} {Interfacial electron-phonon coupling constants extracted
  from intrinsic replica bands in monolayer {FeSe/SrTiO$_3$}},\ }\href
  {https://doi.org/10.1103/physrevlett.127.016803} {\bibfield  {journal}
  {\bibinfo  {journal} {Physical Review Letters}\ }\textbf {\bibinfo {volume}
  {127}},\ \bibinfo {pages} {016803} (\bibinfo {year} {2021})}\BibitemShut
  {NoStop}%
\bibitem [{\citenamefont {Rademaker}\ \emph {et~al.}(2021)\citenamefont
  {Rademaker}, \citenamefont {Alvarez-Suchini}, \citenamefont {Nakatsukasa},
  \citenamefont {Wang},\ and\ \citenamefont {Johnston}}]{Rademaker2021}%
  \BibitemOpen
  \bibfield  {author} {\bibinfo {author} {\bibfnamefont {L.}~\bibnamefont
  {Rademaker}}, \bibinfo {author} {\bibfnamefont {G.}~\bibnamefont
  {Alvarez-Suchini}}, \bibinfo {author} {\bibfnamefont {K.}~\bibnamefont
  {Nakatsukasa}}, \bibinfo {author} {\bibfnamefont {Y.}~\bibnamefont {Wang}},\
  and\ \bibinfo {author} {\bibfnamefont {S.}~\bibnamefont {Johnston}},\
  }\bibfield  {title} {\bibinfo {title} {Enhanced superconductivity in
  {FeSe/SrTiO$_3$} from the combination of forward scattering phonons and spin
  fluctuations},\ }\href {https://doi.org/10.1103/physrevb.103.144504}
  {\bibfield  {journal} {\bibinfo  {journal} {Physical Review B}\ }\textbf
  {\bibinfo {volume} {103}},\ \bibinfo {pages} {144504} (\bibinfo {year}
  {2021})}\BibitemShut {NoStop}%
\bibitem [{\citenamefont {Liu}\ \emph {et~al.}(2021)\citenamefont {Liu},
  \citenamefont {Day}, \citenamefont {Li}, \citenamefont {Roemer},
  \citenamefont {Zhdanovich}, \citenamefont {Gorovikov}, \citenamefont
  {Pedersen}, \citenamefont {Jiang}, \citenamefont {Lee}, \citenamefont
  {Schneider}, \citenamefont {Wong}, \citenamefont {Dosanjh}, \citenamefont
  {Walker}, \citenamefont {Ahn}, \citenamefont {Levy}, \citenamefont
  {Damascelli}, \citenamefont {Sawatzky},\ and\ \citenamefont {Zou}}]{Liu2021}%
  \BibitemOpen
  \bibfield  {author} {\bibinfo {author} {\bibfnamefont {C.}~\bibnamefont
  {Liu}}, \bibinfo {author} {\bibfnamefont {R.~P.}\ \bibnamefont {Day}},
  \bibinfo {author} {\bibfnamefont {F.}~\bibnamefont {Li}}, \bibinfo {author}
  {\bibfnamefont {R.~L.}\ \bibnamefont {Roemer}}, \bibinfo {author}
  {\bibfnamefont {S.}~\bibnamefont {Zhdanovich}}, \bibinfo {author}
  {\bibfnamefont {S.}~\bibnamefont {Gorovikov}}, \bibinfo {author}
  {\bibfnamefont {T.~M.}\ \bibnamefont {Pedersen}}, \bibinfo {author}
  {\bibfnamefont {J.}~\bibnamefont {Jiang}}, \bibinfo {author} {\bibfnamefont
  {S.}~\bibnamefont {Lee}}, \bibinfo {author} {\bibfnamefont {M.}~\bibnamefont
  {Schneider}}, \bibinfo {author} {\bibfnamefont {D.}~\bibnamefont {Wong}},
  \bibinfo {author} {\bibfnamefont {P.}~\bibnamefont {Dosanjh}}, \bibinfo
  {author} {\bibfnamefont {F.~J.}\ \bibnamefont {Walker}}, \bibinfo {author}
  {\bibfnamefont {C.~H.}\ \bibnamefont {Ahn}}, \bibinfo {author} {\bibfnamefont
  {G.}~\bibnamefont {Levy}}, \bibinfo {author} {\bibfnamefont {A.}~\bibnamefont
  {Damascelli}}, \bibinfo {author} {\bibfnamefont {G.~A.}\ \bibnamefont
  {Sawatzky}},\ and\ \bibinfo {author} {\bibfnamefont {K.}~\bibnamefont
  {Zou}},\ }\bibfield  {title} {\bibinfo {title} {High-order replica bands in
  monolayer {FeSe}/{SrTiO$_3$} revealed by polarization-dependent photoemission
  spectroscopy},\ }\bibfield  {journal} {\bibinfo  {journal} {Nature
  Communications}\ }\textbf {\bibinfo {volume} {12}},\ \href
  {https://doi.org/10.1038/s41467-021-24783-5} {10.1038/s41467-021-24783-5}
  (\bibinfo {year} {2021})\BibitemShut {NoStop}%
\bibitem [{\citenamefont {Li}\ and\ \citenamefont {Sawatzky}(2018)}]{Li2018}%
  \BibitemOpen
  \bibfield  {author} {\bibinfo {author} {\bibfnamefont {F.}~\bibnamefont
  {Li}}\ and\ \bibinfo {author} {\bibfnamefont {G.~A.}\ \bibnamefont
  {Sawatzky}},\ }\bibfield  {title} {\bibinfo {title} {Electron phonon coupling
  versus photoelectron energy loss at the origin of replica bands in
  photoemission of {FeSe} on {SrTiO$_3$}},\ }\href
  {https://doi.org/10.1103/physrevlett.120.237001} {\bibfield  {journal}
  {\bibinfo  {journal} {Physical Review Letters}\ }\textbf {\bibinfo {volume}
  {120}},\ \bibinfo {pages} {237001} (\bibinfo {year} {2018})}\BibitemShut
  {NoStop}%
\bibitem [{\citenamefont {Qian}\ \emph {et~al.}(2011)\citenamefont {Qian},
  \citenamefont {Wang}, \citenamefont {Jin}, \citenamefont {Zhang},
  \citenamefont {Richard}, \citenamefont {Xu}, \citenamefont {Dai},
  \citenamefont {Fang}, \citenamefont {Guo}, \citenamefont {Chen},\ and\
  \citenamefont {Ding}}]{Qian2011}%
  \BibitemOpen
  \bibfield  {author} {\bibinfo {author} {\bibfnamefont {T.}~\bibnamefont
  {Qian}}, \bibinfo {author} {\bibfnamefont {X.-P.}\ \bibnamefont {Wang}},
  \bibinfo {author} {\bibfnamefont {W.-C.}\ \bibnamefont {Jin}}, \bibinfo
  {author} {\bibfnamefont {P.}~\bibnamefont {Zhang}}, \bibinfo {author}
  {\bibfnamefont {P.}~\bibnamefont {Richard}}, \bibinfo {author} {\bibfnamefont
  {G.}~\bibnamefont {Xu}}, \bibinfo {author} {\bibfnamefont {X.}~\bibnamefont
  {Dai}}, \bibinfo {author} {\bibfnamefont {Z.}~\bibnamefont {Fang}}, \bibinfo
  {author} {\bibfnamefont {J.-G.}\ \bibnamefont {Guo}}, \bibinfo {author}
  {\bibfnamefont {X.-L.}\ \bibnamefont {Chen}},\ and\ \bibinfo {author}
  {\bibfnamefont {H.}~\bibnamefont {Ding}},\ }\bibfield  {title} {\bibinfo
  {title} {Absence of a holelike fermi surface for the iron-based {KFeSe}
  superconductor revealed by angle-resolved photoemission spectroscopy},\
  }\href {https://doi.org/10.1103/physrevlett.106.187001} {\bibfield  {journal}
  {\bibinfo  {journal} {Physical Review Letters}\ }\textbf {\bibinfo {volume}
  {106}},\ \bibinfo {pages} {187001} (\bibinfo {year} {2011})}\BibitemShut
  {NoStop}%
\bibitem [{\citenamefont {Zhang}\ \emph {et~al.}(2011)\citenamefont {Zhang},
  \citenamefont {Yang}, \citenamefont {Xu}, \citenamefont {Ye}, \citenamefont
  {Chen}, \citenamefont {He}, \citenamefont {Xu}, \citenamefont {Jiang},
  \citenamefont {Xie}, \citenamefont {Ying}, \citenamefont {Wang},
  \citenamefont {Chen}, \citenamefont {Hu}, \citenamefont {Matsunami},
  \citenamefont {Kimura},\ and\ \citenamefont {Feng}}]{Zhang2011}%
  \BibitemOpen
  \bibfield  {author} {\bibinfo {author} {\bibfnamefont {Y.}~\bibnamefont
  {Zhang}}, \bibinfo {author} {\bibfnamefont {L.~X.}\ \bibnamefont {Yang}},
  \bibinfo {author} {\bibfnamefont {M.}~\bibnamefont {Xu}}, \bibinfo {author}
  {\bibfnamefont {Z.~R.}\ \bibnamefont {Ye}}, \bibinfo {author} {\bibfnamefont
  {F.}~\bibnamefont {Chen}}, \bibinfo {author} {\bibfnamefont {C.}~\bibnamefont
  {He}}, \bibinfo {author} {\bibfnamefont {H.~C.}\ \bibnamefont {Xu}}, \bibinfo
  {author} {\bibfnamefont {J.}~\bibnamefont {Jiang}}, \bibinfo {author}
  {\bibfnamefont {B.~P.}\ \bibnamefont {Xie}}, \bibinfo {author} {\bibfnamefont
  {J.~J.}\ \bibnamefont {Ying}}, \bibinfo {author} {\bibfnamefont {X.~F.}\
  \bibnamefont {Wang}}, \bibinfo {author} {\bibfnamefont {X.~H.}\ \bibnamefont
  {Chen}}, \bibinfo {author} {\bibfnamefont {J.~P.}\ \bibnamefont {Hu}},
  \bibinfo {author} {\bibfnamefont {M.}~\bibnamefont {Matsunami}}, \bibinfo
  {author} {\bibfnamefont {S.}~\bibnamefont {Kimura}},\ and\ \bibinfo {author}
  {\bibfnamefont {D.~L.}\ \bibnamefont {Feng}},\ }\bibfield  {title} {\bibinfo
  {title} {Nodeless superconducting gap in {AxFe}2se2 (a=k,cs) revealed by
  angle-resolved photoemission spectroscopy},\ }\href
  {https://doi.org/10.1038/nmat2981} {\bibfield  {journal} {\bibinfo  {journal}
  {Nature Materials}\ }\textbf {\bibinfo {volume} {10}},\ \bibinfo {pages}
  {273} (\bibinfo {year} {2011})}\BibitemShut {NoStop}%
\bibitem [{\citenamefont {Miyata}\ \emph {et~al.}(2015)\citenamefont {Miyata},
  \citenamefont {Nakayama}, \citenamefont {Sugawara}, \citenamefont {Sato},\
  and\ \citenamefont {Takahashi}}]{Miyata2015}%
  \BibitemOpen
  \bibfield  {author} {\bibinfo {author} {\bibfnamefont {Y.}~\bibnamefont
  {Miyata}}, \bibinfo {author} {\bibfnamefont {K.}~\bibnamefont {Nakayama}},
  \bibinfo {author} {\bibfnamefont {K.}~\bibnamefont {Sugawara}}, \bibinfo
  {author} {\bibfnamefont {T.}~\bibnamefont {Sato}},\ and\ \bibinfo {author}
  {\bibfnamefont {T.}~\bibnamefont {Takahashi}},\ }\bibfield  {title} {\bibinfo
  {title} {High-temperature superconductivity in potassium-coated multilayer
  {FeSe} thin films},\ }\href {https://doi.org/10.1038/nmat4302} {\bibfield
  {journal} {\bibinfo  {journal} {Nature Materials}\ }\textbf {\bibinfo
  {volume} {14}},\ \bibinfo {pages} {775} (\bibinfo {year} {2015})}\BibitemShut
  {NoStop}%
\bibitem [{\citenamefont {Lei}\ \emph {et~al.}(2016)\citenamefont {Lei},
  \citenamefont {Cui}, \citenamefont {Xiang}, \citenamefont {Shang},
  \citenamefont {Wang}, \citenamefont {Ye}, \citenamefont {Luo}, \citenamefont
  {Wu}, \citenamefont {Sun},\ and\ \citenamefont {Chen}}]{Lei2016}%
  \BibitemOpen
  \bibfield  {author} {\bibinfo {author} {\bibfnamefont {B.}~\bibnamefont
  {Lei}}, \bibinfo {author} {\bibfnamefont {J.}~\bibnamefont {Cui}}, \bibinfo
  {author} {\bibfnamefont {Z.}~\bibnamefont {Xiang}}, \bibinfo {author}
  {\bibfnamefont {C.}~\bibnamefont {Shang}}, \bibinfo {author} {\bibfnamefont
  {N.}~\bibnamefont {Wang}}, \bibinfo {author} {\bibfnamefont {G.}~\bibnamefont
  {Ye}}, \bibinfo {author} {\bibfnamefont {X.}~\bibnamefont {Luo}}, \bibinfo
  {author} {\bibfnamefont {T.}~\bibnamefont {Wu}}, \bibinfo {author}
  {\bibfnamefont {Z.}~\bibnamefont {Sun}},\ and\ \bibinfo {author}
  {\bibfnamefont {X.}~\bibnamefont {Chen}},\ }\bibfield  {title} {\bibinfo
  {title} {Evolution of high-temperature superconductivity from a low-{T$_c$}
  phase tuned by carrier concentration in {FeSe} thin flakes},\ }\href
  {https://doi.org/10.1103/physrevlett.116.077002} {\bibfield  {journal}
  {\bibinfo  {journal} {Physical Review Letters}\ }\textbf {\bibinfo {volume}
  {116}},\ \bibinfo {pages} {077002} (\bibinfo {year} {2016})}\BibitemShut
  {NoStop}%
\bibitem [{\citenamefont {Stengel}(2011)}]{Stengel2011}%
  \BibitemOpen
  \bibfield  {author} {\bibinfo {author} {\bibfnamefont {M.}~\bibnamefont
  {Stengel}},\ }\bibfield  {title} {\bibinfo {title} {First-principles modeling
  of electrostatically doped perovskite systems},\ }\href
  {https://doi.org/10.1103/physrevlett.106.136803} {\bibfield  {journal}
  {\bibinfo  {journal} {Physical Review Letters}\ }\textbf {\bibinfo {volume}
  {106}},\ \bibinfo {pages} {136803} (\bibinfo {year} {2011})}\BibitemShut
  {NoStop}%
\bibitem [{\citenamefont {Reich}\ \emph {et~al.}(2015)\citenamefont {Reich},
  \citenamefont {Schecter},\ and\ \citenamefont {Shklovskii}}]{Reich2015}%
  \BibitemOpen
  \bibfield  {author} {\bibinfo {author} {\bibfnamefont {K.~V.}\ \bibnamefont
  {Reich}}, \bibinfo {author} {\bibfnamefont {M.}~\bibnamefont {Schecter}},\
  and\ \bibinfo {author} {\bibfnamefont {B.~I.}\ \bibnamefont {Shklovskii}},\
  }\bibfield  {title} {\bibinfo {title} {Accumulation, inversion, and depletion
  layers in {SrTiO$_3$}},\ }\href {https://doi.org/10.1103/physrevb.91.115303}
  {\bibfield  {journal} {\bibinfo  {journal} {Physical Review B}\ }\textbf
  {\bibinfo {volume} {91}},\ \bibinfo {pages} {115303} (\bibinfo {year}
  {2015})}\BibitemShut {NoStop}%
\bibitem [{\citenamefont {Wang}\ \emph
  {et~al.}(2016{\natexlab{b}})\citenamefont {Wang}, \citenamefont {Shen},
  \citenamefont {Pan}, \citenamefont {Hao}, \citenamefont {Ma}, \citenamefont
  {Zhou}, \citenamefont {Steffens}, \citenamefont {Schmalzl}, \citenamefont
  {Forrest}, \citenamefont {Abdel-Hafiez} \emph {et~al.}}]{Wang2015}%
  \BibitemOpen
  \bibfield  {author} {\bibinfo {author} {\bibfnamefont {Q.}~\bibnamefont
  {Wang}}, \bibinfo {author} {\bibfnamefont {Y.}~\bibnamefont {Shen}}, \bibinfo
  {author} {\bibfnamefont {B.}~\bibnamefont {Pan}}, \bibinfo {author}
  {\bibfnamefont {Y.}~\bibnamefont {Hao}}, \bibinfo {author} {\bibfnamefont
  {M.}~\bibnamefont {Ma}}, \bibinfo {author} {\bibfnamefont {F.}~\bibnamefont
  {Zhou}}, \bibinfo {author} {\bibfnamefont {P.}~\bibnamefont {Steffens}},
  \bibinfo {author} {\bibfnamefont {K.}~\bibnamefont {Schmalzl}}, \bibinfo
  {author} {\bibfnamefont {T.}~\bibnamefont {Forrest}}, \bibinfo {author}
  {\bibfnamefont {M.}~\bibnamefont {Abdel-Hafiez}}, \emph {et~al.},\ }\bibfield
   {title} {\bibinfo {title} {Strong interplay between stripe spin
  fluctuations, nematicity and superconductivity in {FeSe}},\ }\href
  {https://doi.org/10.1038/nmat4492} {\bibfield  {journal} {\bibinfo  {journal}
  {Nature Materials}\ }\textbf {\bibinfo {volume} {15}},\ \bibinfo {pages}
  {159} (\bibinfo {year} {2016}{\natexlab{b}})}\BibitemShut {NoStop}%
\bibitem [{\citenamefont {Chen}\ \emph {et~al.}(2019)\citenamefont {Chen},
  \citenamefont {Chen}, \citenamefont {Kreisel}, \citenamefont {Lu},
  \citenamefont {Schneidewind}, \citenamefont {Qiu}, \citenamefont {Park},
  \citenamefont {Perring}, \citenamefont {Stewart}, \citenamefont {Cao},
  \citenamefont {Zhang}, \citenamefont {Li}, \citenamefont {Rong},
  \citenamefont {Wei}, \citenamefont {Andersen}, \citenamefont {Hirschfeld},
  \citenamefont {Broholm},\ and\ \citenamefont {Dai}}]{Chen2019}%
  \BibitemOpen
  \bibfield  {author} {\bibinfo {author} {\bibfnamefont {T.}~\bibnamefont
  {Chen}}, \bibinfo {author} {\bibfnamefont {Y.}~\bibnamefont {Chen}}, \bibinfo
  {author} {\bibfnamefont {A.}~\bibnamefont {Kreisel}}, \bibinfo {author}
  {\bibfnamefont {X.}~\bibnamefont {Lu}}, \bibinfo {author} {\bibfnamefont
  {A.}~\bibnamefont {Schneidewind}}, \bibinfo {author} {\bibfnamefont
  {Y.}~\bibnamefont {Qiu}}, \bibinfo {author} {\bibfnamefont {J.~T.}\
  \bibnamefont {Park}}, \bibinfo {author} {\bibfnamefont {T.~G.}\ \bibnamefont
  {Perring}}, \bibinfo {author} {\bibfnamefont {J.~R.}\ \bibnamefont
  {Stewart}}, \bibinfo {author} {\bibfnamefont {H.}~\bibnamefont {Cao}},
  \bibinfo {author} {\bibfnamefont {R.}~\bibnamefont {Zhang}}, \bibinfo
  {author} {\bibfnamefont {Y.}~\bibnamefont {Li}}, \bibinfo {author}
  {\bibfnamefont {Y.}~\bibnamefont {Rong}}, \bibinfo {author} {\bibfnamefont
  {Y.}~\bibnamefont {Wei}}, \bibinfo {author} {\bibfnamefont {B.~M.}\
  \bibnamefont {Andersen}}, \bibinfo {author} {\bibfnamefont {P.~J.}\
  \bibnamefont {Hirschfeld}}, \bibinfo {author} {\bibfnamefont
  {C.}~\bibnamefont {Broholm}},\ and\ \bibinfo {author} {\bibfnamefont
  {P.}~\bibnamefont {Dai}},\ }\bibfield  {title} {\bibinfo {title} {Anisotropic
  spin fluctuations in detwinned {FeSe}},\ }\href
  {https://doi.org/10.1038/s41563-019-0369-5} {\bibfield  {journal} {\bibinfo
  {journal} {Nature Materials}\ }\textbf {\bibinfo {volume} {18}},\ \bibinfo
  {pages} {709} (\bibinfo {year} {2019})}\BibitemShut {NoStop}%
\bibitem [{\citenamefont {Phelan}\ \emph {et~al.}(2009)\citenamefont {Phelan},
  \citenamefont {Millican}, \citenamefont {Thomas}, \citenamefont {Le\~ao},
  \citenamefont {Qiu},\ and\ \citenamefont {Paul}}]{Phelan2009}%
  \BibitemOpen
  \bibfield  {author} {\bibinfo {author} {\bibfnamefont {D.}~\bibnamefont
  {Phelan}}, \bibinfo {author} {\bibfnamefont {J.~N.}\ \bibnamefont
  {Millican}}, \bibinfo {author} {\bibfnamefont {E.~L.}\ \bibnamefont
  {Thomas}}, \bibinfo {author} {\bibfnamefont {J.~B.}\ \bibnamefont {Le\~ao}},
  \bibinfo {author} {\bibfnamefont {Y.}~\bibnamefont {Qiu}},\ and\ \bibinfo
  {author} {\bibfnamefont {R.}~\bibnamefont {Paul}},\ }\bibfield  {title}
  {\bibinfo {title} {Neutron scattering measurements of the phonon density of
  states of {FeSe$_{1-x}$} superconductors},\ }\href
  {https://doi.org/10.1103/PhysRevB.79.014519} {\bibfield  {journal} {\bibinfo
  {journal} {Phys. Rev. B}\ }\textbf {\bibinfo {volume} {79}},\ \bibinfo
  {pages} {014519} (\bibinfo {year} {2009})}\BibitemShut {NoStop}%
\bibitem [{\citenamefont {Ye}\ \emph {et~al.}(2013)\citenamefont {Ye},
  \citenamefont {Liu},\ and\ \citenamefont {Lu}}]{Ye2013}%
  \BibitemOpen
  \bibfield  {author} {\bibinfo {author} {\bibfnamefont {Q.-Q.}\ \bibnamefont
  {Ye}}, \bibinfo {author} {\bibfnamefont {K.}~\bibnamefont {Liu}},\ and\
  \bibinfo {author} {\bibfnamefont {Z.-Y.}\ \bibnamefont {Lu}},\ }\bibfield
  {title} {\bibinfo {title} {Influence of spin-phonon coupling on
  antiferromagnetic spin fluctuations in fese under pressure: First-principles
  calculations with van der waals corrections},\ }\href
  {https://doi.org/10.1103/PhysRevB.88.205130} {\bibfield  {journal} {\bibinfo
  {journal} {Phys. Rev. B}\ }\textbf {\bibinfo {volume} {88}},\ \bibinfo
  {pages} {205130} (\bibinfo {year} {2013})}\BibitemShut {NoStop}%
\bibitem [{\citenamefont {Mazin}(2015)}]{Mazin2015}%
  \BibitemOpen
  \bibfield  {author} {\bibinfo {author} {\bibfnamefont {I.~I.}\ \bibnamefont
  {Mazin}},\ }\bibfield  {title} {\bibinfo {title} {Superconductivity: The fese
  riddle},\ }\href {http://dx.doi.org/10.1038/nmat4371} {\bibfield  {journal}
  {\bibinfo  {journal} {Nat Mater}\ }\textbf {\bibinfo {volume} {14}},\
  \bibinfo {pages} {755} (\bibinfo {year} {2015})}\BibitemShut {NoStop}%
\bibitem [{\citenamefont {Schrodi}\ \emph
  {et~al.}(2020{\natexlab{a}})\citenamefont {Schrodi}, \citenamefont {Aperis},\
  and\ \citenamefont {Oppeneer}}]{Schrodi2020}%
  \BibitemOpen
  \bibfield  {author} {\bibinfo {author} {\bibfnamefont {F.}~\bibnamefont
  {Schrodi}}, \bibinfo {author} {\bibfnamefont {A.}~\bibnamefont {Aperis}},\
  and\ \bibinfo {author} {\bibfnamefont {P.~M.}\ \bibnamefont {Oppeneer}},\
  }\bibfield  {title} {\bibinfo {title} {Eliashberg theory for spin fluctuation
  mediated superconductivity: Application to bulk and monolayer {FeSe}},\
  }\href {https://doi.org/10.1103/physrevb.102.014502} {\bibfield  {journal}
  {\bibinfo  {journal} {Physical Review B}\ }\textbf {\bibinfo {volume}
  {102}},\ \bibinfo {pages} {014502} (\bibinfo {year}
  {2020}{\natexlab{a}})}\BibitemShut {NoStop}%
\bibitem [{\citenamefont {Acharya}\ \emph {et~al.}(2021)\citenamefont
  {Acharya}, \citenamefont {Pashov}, \citenamefont {Jamet},\ and\ \citenamefont
  {van Schilfgaarde}}]{Acharya2021}%
  \BibitemOpen
  \bibfield  {author} {\bibinfo {author} {\bibfnamefont {S.}~\bibnamefont
  {Acharya}}, \bibinfo {author} {\bibfnamefont {D.}~\bibnamefont {Pashov}},
  \bibinfo {author} {\bibfnamefont {F.}~\bibnamefont {Jamet}},\ and\ \bibinfo
  {author} {\bibfnamefont {M.}~\bibnamefont {van Schilfgaarde}},\ }\bibfield
  {title} {\bibinfo {title} {Electronic origin of tc in bulk and monolayer
  {FeSe}},\ }\href {https://doi.org/10.3390/sym13020169} {\bibfield  {journal}
  {\bibinfo  {journal} {Symmetry}\ }\textbf {\bibinfo {volume} {13}},\ \bibinfo
  {pages} {169} (\bibinfo {year} {2021})}\BibitemShut {NoStop}%
\bibitem [{\citenamefont {Pelliciari}\ \emph {et~al.}(2021)\citenamefont
  {Pelliciari}, \citenamefont {Karakuzu}, \citenamefont {Song}, \citenamefont
  {Arpaia}, \citenamefont {Nag}, \citenamefont {Rossi}, \citenamefont {Li},
  \citenamefont {Yu}, \citenamefont {Chen}, \citenamefont {Peng}, \citenamefont
  {Garc{\'{\i}}a-Fern{\'{a}}ndez}, \citenamefont {Walters}, \citenamefont
  {Wang}, \citenamefont {Zhao}, \citenamefont {Ghiringhelli}, \citenamefont
  {Feng}, \citenamefont {Maier}, \citenamefont {Zhou}, \citenamefont
  {Johnston},\ and\ \citenamefont {Comin}}]{Pelliciari2021}%
  \BibitemOpen
  \bibfield  {author} {\bibinfo {author} {\bibfnamefont {J.}~\bibnamefont
  {Pelliciari}}, \bibinfo {author} {\bibfnamefont {S.}~\bibnamefont
  {Karakuzu}}, \bibinfo {author} {\bibfnamefont {Q.}~\bibnamefont {Song}},
  \bibinfo {author} {\bibfnamefont {R.}~\bibnamefont {Arpaia}}, \bibinfo
  {author} {\bibfnamefont {A.}~\bibnamefont {Nag}}, \bibinfo {author}
  {\bibfnamefont {M.}~\bibnamefont {Rossi}}, \bibinfo {author} {\bibfnamefont
  {J.}~\bibnamefont {Li}}, \bibinfo {author} {\bibfnamefont {T.}~\bibnamefont
  {Yu}}, \bibinfo {author} {\bibfnamefont {X.}~\bibnamefont {Chen}}, \bibinfo
  {author} {\bibfnamefont {R.}~\bibnamefont {Peng}}, \bibinfo {author}
  {\bibfnamefont {M.}~\bibnamefont {Garc{\'{\i}}a-Fern{\'{a}}ndez}}, \bibinfo
  {author} {\bibfnamefont {A.~C.}\ \bibnamefont {Walters}}, \bibinfo {author}
  {\bibfnamefont {Q.}~\bibnamefont {Wang}}, \bibinfo {author} {\bibfnamefont
  {J.}~\bibnamefont {Zhao}}, \bibinfo {author} {\bibfnamefont {G.}~\bibnamefont
  {Ghiringhelli}}, \bibinfo {author} {\bibfnamefont {D.}~\bibnamefont {Feng}},
  \bibinfo {author} {\bibfnamefont {T.~A.}\ \bibnamefont {Maier}}, \bibinfo
  {author} {\bibfnamefont {K.-J.}\ \bibnamefont {Zhou}}, \bibinfo {author}
  {\bibfnamefont {S.}~\bibnamefont {Johnston}},\ and\ \bibinfo {author}
  {\bibfnamefont {R.}~\bibnamefont {Comin}},\ }\bibfield  {title} {\bibinfo
  {title} {Evolution of spin excitations from bulk to monolayer {FeSe}},\
  }\bibfield  {journal} {\bibinfo  {journal} {Nature Communications}\ }\textbf
  {\bibinfo {volume} {12}},\ \href {https://doi.org/10.1038/s41467-021-23317-3}
  {10.1038/s41467-021-23317-3} (\bibinfo {year} {2021})\BibitemShut {NoStop}%
\bibitem [{\citenamefont {Graser}\ \emph {et~al.}(2009)\citenamefont {Graser},
  \citenamefont {Maier}, \citenamefont {Hirschfeld},\ and\ \citenamefont
  {Scalapino}}]{Graser2009}%
  \BibitemOpen
  \bibfield  {author} {\bibinfo {author} {\bibfnamefont {S.}~\bibnamefont
  {Graser}}, \bibinfo {author} {\bibfnamefont {T.~A.}\ \bibnamefont {Maier}},
  \bibinfo {author} {\bibfnamefont {P.~J.}\ \bibnamefont {Hirschfeld}},\ and\
  \bibinfo {author} {\bibfnamefont {D.~J.}\ \bibnamefont {Scalapino}},\
  }\bibfield  {title} {\bibinfo {title} {Near-degeneracy of several pairing
  channels in multiorbital models for the {Fe} pnictides},\ }\href
  {https://doi.org/10.1088/1367-2630/11/2/025016} {\bibfield  {journal}
  {\bibinfo  {journal} {New Journal of Physics}\ }\textbf {\bibinfo {volume}
  {11}},\ \bibinfo {pages} {025016} (\bibinfo {year} {2009})}\BibitemShut
  {NoStop}%
\bibitem [{\citenamefont {Linscheid}\ \emph {et~al.}(2016)\citenamefont
  {Linscheid}, \citenamefont {Maiti}, \citenamefont {Wang}, \citenamefont
  {Johnston},\ and\ \citenamefont {Hirschfeld}}]{Linscheid2016}%
  \BibitemOpen
  \bibfield  {author} {\bibinfo {author} {\bibfnamefont {A.}~\bibnamefont
  {Linscheid}}, \bibinfo {author} {\bibfnamefont {S.}~\bibnamefont {Maiti}},
  \bibinfo {author} {\bibfnamefont {Y.}~\bibnamefont {Wang}}, \bibinfo {author}
  {\bibfnamefont {S.}~\bibnamefont {Johnston}},\ and\ \bibinfo {author}
  {\bibfnamefont {P.}~\bibnamefont {Hirschfeld}},\ }\bibfield  {title}
  {\bibinfo {title} {High {T$_{C}$} via spin fluctuations from incipient bands:
  Application to monolayers and intercalates of {FeSe}},\ }\href
  {https://doi.org/10.1103/physrevlett.117.077003} {\bibfield  {journal}
  {\bibinfo  {journal} {Physical Review Letters}\ }\textbf {\bibinfo {volume}
  {117}},\ \bibinfo {pages} {077003} (\bibinfo {year} {2016})}\BibitemShut
  {NoStop}%
\bibitem [{\citenamefont {Liu}\ \emph {et~al.}(2019)\citenamefont {Liu},
  \citenamefont {Wang}, \citenamefont {Ye}, \citenamefont {Chen}, \citenamefont
  {Liu}, \citenamefont {Wang}, \citenamefont {Wang},\ and\ \citenamefont
  {Wang}}]{Liu2019}%
  \BibitemOpen
  \bibfield  {author} {\bibinfo {author} {\bibfnamefont {C.}~\bibnamefont
  {Liu}}, \bibinfo {author} {\bibfnamefont {Z.}~\bibnamefont {Wang}}, \bibinfo
  {author} {\bibfnamefont {S.}~\bibnamefont {Ye}}, \bibinfo {author}
  {\bibfnamefont {C.}~\bibnamefont {Chen}}, \bibinfo {author} {\bibfnamefont
  {Y.}~\bibnamefont {Liu}}, \bibinfo {author} {\bibfnamefont {Q.}~\bibnamefont
  {Wang}}, \bibinfo {author} {\bibfnamefont {Q.-H.}\ \bibnamefont {Wang}},\
  and\ \bibinfo {author} {\bibfnamefont {J.}~\bibnamefont {Wang}},\ }\bibfield
  {title} {\bibinfo {title} {Detection of bosonic mode as a signature of
  magnetic excitation in one-unit-cell {FeSe} on {SrTiO$_3$}},\ }\href
  {https://doi.org/10.1021/acs.nanolett.9b00144} {\bibfield  {journal}
  {\bibinfo  {journal} {Nano Letters}\ }\textbf {\bibinfo {volume} {19}},\
  \bibinfo {pages} {3464} (\bibinfo {year} {2019})}\BibitemShut {NoStop}%
\bibitem [{\citenamefont {Schrodi}\ \emph
  {et~al.}(2020{\natexlab{b}})\citenamefont {Schrodi}, \citenamefont {Aperis},\
  and\ \citenamefont {Oppeneer}}]{Schrodi2020a}%
  \BibitemOpen
  \bibfield  {author} {\bibinfo {author} {\bibfnamefont {F.}~\bibnamefont
  {Schrodi}}, \bibinfo {author} {\bibfnamefont {A.}~\bibnamefont {Aperis}},\
  and\ \bibinfo {author} {\bibfnamefont {P.~M.}\ \bibnamefont {Oppeneer}},\
  }\bibfield  {title} {\bibinfo {title} {Multichannel superconductivity of
  monolayer {FeSe} on {SrTiO}$_3$: Interplay of spin fluctuations and
  electron-phonon interaction},\ }\href
  {https://doi.org/10.1103/physrevb.102.180501} {\bibfield  {journal} {\bibinfo
   {journal} {Physical Review B}\ }\textbf {\bibinfo {volume} {102}},\ \bibinfo
  {pages} {180501} (\bibinfo {year} {2020}{\natexlab{b}})}\BibitemShut
  {NoStop}%
\bibitem [{\citenamefont {Li}\ \emph {et~al.}(2014)\citenamefont {Li},
  \citenamefont {Peng}, \citenamefont {Zhang}, \citenamefont {Zhang},
  \citenamefont {Ding}, \citenamefont {Deng}, \citenamefont {Chang},
  \citenamefont {Song}, \citenamefont {Ji}, \citenamefont {Wang}, \citenamefont
  {He}, \citenamefont {Chen}, \citenamefont {Xue},\ and\ \citenamefont
  {Ma}}]{Li2014}%
  \BibitemOpen
  \bibfield  {author} {\bibinfo {author} {\bibfnamefont {Z.}~\bibnamefont
  {Li}}, \bibinfo {author} {\bibfnamefont {J.-P.}\ \bibnamefont {Peng}},
  \bibinfo {author} {\bibfnamefont {H.-M.}\ \bibnamefont {Zhang}}, \bibinfo
  {author} {\bibfnamefont {W.-H.}\ \bibnamefont {Zhang}}, \bibinfo {author}
  {\bibfnamefont {H.}~\bibnamefont {Ding}}, \bibinfo {author} {\bibfnamefont
  {P.}~\bibnamefont {Deng}}, \bibinfo {author} {\bibfnamefont {K.}~\bibnamefont
  {Chang}}, \bibinfo {author} {\bibfnamefont {C.-L.}\ \bibnamefont {Song}},
  \bibinfo {author} {\bibfnamefont {S.-H.}\ \bibnamefont {Ji}}, \bibinfo
  {author} {\bibfnamefont {L.}~\bibnamefont {Wang}}, \bibinfo {author}
  {\bibfnamefont {K.}~\bibnamefont {He}}, \bibinfo {author} {\bibfnamefont
  {X.}~\bibnamefont {Chen}}, \bibinfo {author} {\bibfnamefont {Q.-K.}\
  \bibnamefont {Xue}},\ and\ \bibinfo {author} {\bibfnamefont {X.-C.}\
  \bibnamefont {Ma}},\ }\bibfield  {title} {\bibinfo {title} {Molecular beam
  epitaxy growth and post-growth annealing of {FeSe} films on {SrTiO$_3$}: a
  scanning tunneling microscopy study},\ }\href
  {https://doi.org/10.1088/0953-8984/26/26/265002} {\bibfield  {journal}
  {\bibinfo  {journal} {Journal of Physics: Condensed Matter}\ }\textbf
  {\bibinfo {volume} {26}},\ \bibinfo {pages} {265002} (\bibinfo {year}
  {2014})}\BibitemShut {NoStop}%
\bibitem [{\citenamefont {Ibach}\ \emph {et~al.}(2003)\citenamefont {Ibach},
  \citenamefont {Bruchmann}, \citenamefont {Vollmer}, \citenamefont {Etzkorn},
  \citenamefont {Kumar},\ and\ \citenamefont {Kirschner}}]{Ibach2003}%
  \BibitemOpen
  \bibfield  {author} {\bibinfo {author} {\bibfnamefont {H.}~\bibnamefont
  {Ibach}}, \bibinfo {author} {\bibfnamefont {D.}~\bibnamefont {Bruchmann}},
  \bibinfo {author} {\bibfnamefont {R.}~\bibnamefont {Vollmer}}, \bibinfo
  {author} {\bibfnamefont {M.}~\bibnamefont {Etzkorn}}, \bibinfo {author}
  {\bibfnamefont {P.~S.~A.}\ \bibnamefont {Kumar}},\ and\ \bibinfo {author}
  {\bibfnamefont {J.}~\bibnamefont {Kirschner}},\ }\bibfield  {title} {\bibinfo
  {title} {A novel spectrometer for spin-polarized electron energy-loss
  spectroscopy},\ }\href {https://doi.org/10.1063/1.1597954} {\bibfield
  {journal} {\bibinfo  {journal} {Rev. Sci. Instrum.}\ }\textbf {\bibinfo
  {volume} {74}},\ \bibinfo {pages} {4089} (\bibinfo {year}
  {2003})}\BibitemShut {NoStop}%
\bibitem [{\citenamefont {Vollmer}\ \emph {et~al.}(2003)\citenamefont
  {Vollmer}, \citenamefont {Etzkorn}, \citenamefont {Kumar}, \citenamefont
  {Ibach},\ and\ \citenamefont {Kirschner}}]{Vollmer2003}%
  \BibitemOpen
  \bibfield  {author} {\bibinfo {author} {\bibfnamefont {R.}~\bibnamefont
  {Vollmer}}, \bibinfo {author} {\bibfnamefont {M.}~\bibnamefont {Etzkorn}},
  \bibinfo {author} {\bibfnamefont {P.~S.~A.}\ \bibnamefont {Kumar}}, \bibinfo
  {author} {\bibfnamefont {H.}~\bibnamefont {Ibach}},\ and\ \bibinfo {author}
  {\bibfnamefont {J.}~\bibnamefont {Kirschner}},\ }\bibfield  {title} {\bibinfo
  {title} {Spin-polarized electron energy loss spectroscopy of high energy,
  large wave vector spin waves in ultrathin fcc {C}o films on {C}u(001)},\
  }\href {https://doi.org/10.1103/PhysRevLett.91.147201} {\bibfield  {journal}
  {\bibinfo  {journal} {Phys. Rev. Lett.}\ }\textbf {\bibinfo {volume} {91}},\
  \bibinfo {pages} {147201} (\bibinfo {year} {2003})}\BibitemShut {NoStop}%
\bibitem [{\citenamefont {Zakeri}(2014)}]{Zakeri2014}%
  \BibitemOpen
  \bibfield  {author} {\bibinfo {author} {\bibfnamefont {K.}~\bibnamefont
  {Zakeri}},\ }\bibfield  {title} {\bibinfo {title} {Elementary spin
  excitations in ultrathin itinerant magnets},\ }\href
  {https://doi.org/h10.1016/j.physrep.2014.08.001} {\bibfield  {journal}
  {\bibinfo  {journal} {Phys. Rep.}\ }\textbf {\bibinfo {volume} {545}},\
  \bibinfo {pages} {47} (\bibinfo {year} {2014})}\BibitemShut {NoStop}%
\bibitem [{\citenamefont {Zakeri}\ \emph {et~al.}(2021)\citenamefont {Zakeri},
  \citenamefont {Wettstein},\ and\ \citenamefont {S\"urgers}}]{Zakeri2021}%
  \BibitemOpen
  \bibfield  {author} {\bibinfo {author} {\bibfnamefont {K.}~\bibnamefont
  {Zakeri}}, \bibinfo {author} {\bibfnamefont {J.}~\bibnamefont {Wettstein}},\
  and\ \bibinfo {author} {\bibfnamefont {C.}~\bibnamefont {S\"urgers}},\
  }\bibfield  {title} {\bibinfo {title} {Generation of spin-polarized hot
  electrons at topological insulators surfaces by scattering from collective
  charge excitations},\ }\bibfield  {journal} {\bibinfo  {journal}
  {Communications Physics}\ }\textbf {\bibinfo {volume} {4}},\ \href
  {https://doi.org/10.1038/s42005-021-00729-7} {10.1038/s42005-021-00729-7}
  (\bibinfo {year} {2021})\BibitemShut {NoStop}%
\bibitem [{\citenamefont {\ifmmode \check{S}\else
  \v{S}\fi{}unji\ifmmode~\acute{c}\else \'{c}\fi{}}\ and\ \citenamefont
  {Lucas}(1971)}]{Sunjic1971}%
  \BibitemOpen
  \bibfield  {author} {\bibinfo {author} {\bibfnamefont {M.}~\bibnamefont
  {\ifmmode \check{S}\else \v{S}\fi{}unji\ifmmode~\acute{c}\else \'{c}\fi{}}}\
  and\ \bibinfo {author} {\bibfnamefont {A.~A.}\ \bibnamefont {Lucas}},\
  }\bibfield  {title} {\bibinfo {title} {Multiple plasmon effects in the
  energy-loss spectra of electrons in thin films},\ }\href
  {https://doi.org/10.1103/PhysRevB.3.719} {\bibfield  {journal} {\bibinfo
  {journal} {Phys. Rev. B}\ }\textbf {\bibinfo {volume} {3}},\ \bibinfo {pages}
  {719} (\bibinfo {year} {1971})}\BibitemShut {NoStop}%
\bibitem [{\citenamefont {Lucas}\ and\ \citenamefont
  {{\v{S}}unji{\'{c}}}(1972)}]{Lucas1972}%
  \BibitemOpen
  \bibfield  {author} {\bibinfo {author} {\bibfnamefont {A.~A.}\ \bibnamefont
  {Lucas}}\ and\ \bibinfo {author} {\bibfnamefont {M.}~\bibnamefont
  {{\v{S}}unji{\'{c}}}},\ }\bibfield  {title} {\bibinfo {title} {Fast-electron
  spectroscopy of collective excitations in solids},\ }\href
  {http://www.sciencedirect.com/science/article/pii/0079681672900020}
  {\bibfield  {journal} {\bibinfo  {journal} {Progress in Surface Science}\
  }\textbf {\bibinfo {volume} {2}},\ \bibinfo {pages} {75} (\bibinfo {year}
  {1972})}\BibitemShut {NoStop}%
\bibitem [{\citenamefont {Lambin}\ \emph {et~al.}(1990)\citenamefont {Lambin},
  \citenamefont {Vigneron},\ and\ \citenamefont {Lucas}}]{Lambin1990}%
  \BibitemOpen
  \bibfield  {author} {\bibinfo {author} {\bibfnamefont {P.}~\bibnamefont
  {Lambin}}, \bibinfo {author} {\bibfnamefont {J.-P.}\ \bibnamefont
  {Vigneron}},\ and\ \bibinfo {author} {\bibfnamefont {A.}~\bibnamefont
  {Lucas}},\ }\bibfield  {title} {\bibinfo {title} {Computation of the surface
  electron-energy-loss spectrum in specular geometry for an arbitrary
  plane-stratified medium},\ }\href
  {https://doi.org/10.1016/0010-4655(90)90034-X} {\bibfield  {journal}
  {\bibinfo  {journal} {Computer Physics Communications}\ }\textbf {\bibinfo
  {volume} {60}},\ \bibinfo {pages} {351} (\bibinfo {year} {1990})}\BibitemShut
  {NoStop}%
\bibitem [{\citenamefont {Lazzari}\ \emph {et~al.}(2018)\citenamefont
  {Lazzari}, \citenamefont {Li},\ and\ \citenamefont {Jupille}}]{Lazzari2018}%
  \BibitemOpen
  \bibfield  {author} {\bibinfo {author} {\bibfnamefont {R.}~\bibnamefont
  {Lazzari}}, \bibinfo {author} {\bibfnamefont {J.}~\bibnamefont {Li}},\ and\
  \bibinfo {author} {\bibfnamefont {J.}~\bibnamefont {Jupille}},\ }\bibfield
  {title} {\bibinfo {title} {Dielectric study of the interplay between charge
  carriers and electron energy losses in reduced titanium dioxide},\ }\href
  {https://doi.org/10.1103/PhysRevB.98.075432} {\bibfield  {journal} {\bibinfo
  {journal} {Physical Review B}\ }\textbf {\bibinfo {volume} {98}},\ \bibinfo
  {pages} {075432} (\bibinfo {year} {2018})}\BibitemShut {NoStop}%
\bibitem [{\citenamefont {Yuan}\ \emph {et~al.}(2012)\citenamefont {Yuan},
  \citenamefont {Dong}, \citenamefont {Song}, \citenamefont {Zheng},
  \citenamefont {Chen}, \citenamefont {Hu}, \citenamefont {Li},\ and\
  \citenamefont {Wang}}]{Yuan2012}%
  \BibitemOpen
  \bibfield  {author} {\bibinfo {author} {\bibfnamefont {R.~H.}\ \bibnamefont
  {Yuan}}, \bibinfo {author} {\bibfnamefont {T.}~\bibnamefont {Dong}}, \bibinfo
  {author} {\bibfnamefont {Y.~J.}\ \bibnamefont {Song}}, \bibinfo {author}
  {\bibfnamefont {P.}~\bibnamefont {Zheng}}, \bibinfo {author} {\bibfnamefont
  {G.~F.}\ \bibnamefont {Chen}}, \bibinfo {author} {\bibfnamefont {J.~P.}\
  \bibnamefont {Hu}}, \bibinfo {author} {\bibfnamefont {J.~Q.}\ \bibnamefont
  {Li}},\ and\ \bibinfo {author} {\bibfnamefont {N.~L.}\ \bibnamefont {Wang}},\
  }\bibfield  {title} {\bibinfo {title} {Nanoscale phase separation of
  antiferromagnetic order and superconductivity in
  {K$_{0.75}$Fe$_{1.75}$Se$_2$}},\ }\bibfield  {journal} {\bibinfo  {journal}
  {Scientific Reports}\ }\textbf {\bibinfo {volume} {2}},\ \href
  {https://doi.org/10.1038/srep00221} {10.1038/srep00221} (\bibinfo {year}
  {2012})\BibitemShut {NoStop}%
\bibitem [{\citenamefont {Eagles}\ \emph {et~al.}(1996)\citenamefont {Eagles},
  \citenamefont {Georgiev},\ and\ \citenamefont {Petrova}}]{Eagles1996}%
  \BibitemOpen
  \bibfield  {author} {\bibinfo {author} {\bibfnamefont {D.~M.}\ \bibnamefont
  {Eagles}}, \bibinfo {author} {\bibfnamefont {M.}~\bibnamefont {Georgiev}},\
  and\ \bibinfo {author} {\bibfnamefont {P.~C.}\ \bibnamefont {Petrova}},\
  }\bibfield  {title} {\bibinfo {title} {Explanation for the temperature
  dependence of plasma frequencies in {SrTiO$_3$} using mixed-polaron theory},\
  }\href {https://doi.org/10.1103/physrevb.54.22} {\bibfield  {journal}
  {\bibinfo  {journal} {Physical Review B}\ }\textbf {\bibinfo {volume} {54}},\
  \bibinfo {pages} {22} (\bibinfo {year} {1996})}\BibitemShut {NoStop}%
\end{thebibliography}%


%
\twocolumngrid

\end{document}